\documentclass[aps,pra,reprint,twocolumn,floatfix,amsmath,amssymb]{revtex4-2}
\usepackage{bm}
\usepackage{mathtools}
\usepackage{physics}
\usepackage{graphicx}
\usepackage{etoolbox}
\usepackage[x11names,dvipsnames]{xcolor}
\usepackage[colorlinks=true,citecolor=blue,linkcolor=blue,urlcolor=blue]{hyperref}
\hypersetup{pdftitle={High-Fidelity CZ gate via Measurement-Based Quantum Computation Using Spin-1 XXZ Chains},pdfauthor={Hiroki Ohta and Shunji Tsuchiya}}
\allowdisplaybreaks

\newcommand{\combinedbibpart}{main}
\begin{document}
\title{High-Fidelity $CZ$ gate via Measurement-Based Quantum Computation Using Spin-1 $XXZ$ Chains}
\author{Hiroki Ohta}
\author{Shunji Tsuchiya}
\affiliation{Department of Physics, Chuo University, Bunkyo, Tokyo 112-8551, Japan}

\begin{abstract}
We show that the ground states of spin-1 $XXZ$ chains with single-ion anisotropy $D$ and Ising-like anisotropy $J$ can be used  for  implementation of a $CZ$ gate in measurement-based quantum computation. We consider two spin-1 $XXZ$ chains dynamically coupled via a short interaction as the resource state for the $CZ$ gate and apply the same protocol as that used for the Affleck--Kennedy--Lieb--Tasaki state. The gate fidelity of the $CZ$ gate is evaluated for the dynamically coupled spin-1 $XXZ$ chains and is found to be enhanced by tuning the anisotropies $D$ and $J$ toward the antiferromagnetic phase in the Haldane phase, exceeding the value of $7/12$ obtained for the Affleck--Kennedy--Lieb--Tasaki state. We further derive a compact expression for the gate fidelity of the $CZ$ gate, valid throughout the Haldane phase, in terms of nonlocal correlation functions and the probability that the measurement
fails. This expression reveals that antiferromagnetic correlations enhance the gate fidelity of the $CZ$ gate, suppressing the failure probability.
\end{abstract}
\maketitle
{\it Introduction.}---Quantum computation is a novel paradigm for information processing that leverages the principles of quantum mechanics \cite{Nielsen2011}. The standard form of quantum computation is based on quantum circuits, in which computation is performed by applying a sequence of unitary gates to an initial quantum state \cite{Nielsen2011}.
On the other hand, {\it measurement-based quantum computation} (MBQC) takes a different approach---computation proceeds through single-particle measurements adaptively performed on a preprepared quantum many-body resource state \cite{Raussendorf2001}. Since the computational power of MBQC strongly depends on the resource state, it is important to identify many-body states that can serve as resources for MBQC and what determines their computational power.

The cluster state \cite{Briegel2001,Walther2005,Chen2007} and the Affleck-Kennedy-Lieb-Tasaki (AKLT) state \cite{Affleck1987,Affleck1988} are well-known resource states for MBQC \cite{Raussendorf2001,Brennen2008}, and both are classified as symmetry-protected topological (SPT) phases \cite{Pollman2010,Pollman2012,Chen2013,Wen2017}. Many-body states in an SPT phase cannot be transformed into product states by finite-depth local unitary circuits which preserve the symmetry associated with the SPT phase \cite{Pollman2010,Chen2013,Wen2017}. In particular, the one-dimensional (1D) spin-1 AKLT state belongs to {\it the Haldane phase} \cite{Haldane1983}, an SPT phase protected by the $\mathbb{Z}_2\times\mathbb{Z}_2$ symmetry generated by global $\pi$ rotations \cite{Pollman2012}.

In recent years, the Haldane phase has been experimentally realized on various experimental platforms, such as photonic systems, fermionic cold-atom systems, and superconducting quantum processors \cite{Kaltenbaek2010,Sompet2022,Smith2023}. Furthermore, single-qubit MBQC has been demonstrated using the photonic AKLT state {\cite{Kaltenbaek2010}}. 
Meanwhile, recent theoretical studies have shown that ground states in the Haldane phase can serve as resource states for single-qubit MBQC \cite{Miyake2010,Barlett2010,Dominic2012,Raussendorf2017,Raussendorf2023,yang2024}. In particular, Refs.~\citealp{Raussendorf2023} and \citealp{yang2024} showed that the computational power of many-body states can be characterized by the string order parameter \cite{den1989} of the Haldane phase.

\begin{figure}[htbp]
\centering
\includegraphics[width=0.8\columnwidth]{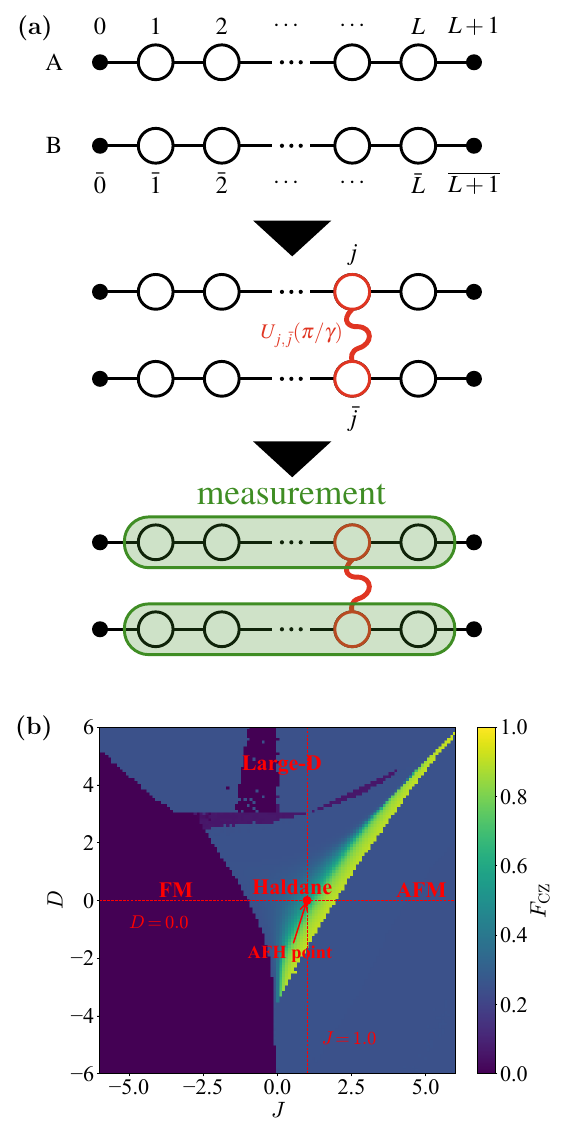}
\caption{(a)~Schematic illustration of the implementation of a $CZ$ gate using two spin-1 chains as the resource state. The black dots and the solid circles represent spin-$1/2$ sites and spin-1 sites, respectively. Here, $L$ is the number of spin-1s, and the labels $0$ and $L+1$ denote the spin-$1/2$s attached at the left and right ends of each chain, which serve as the input and output qubits, respectively. The $CZ$ gate is implemented by dynamically coupling two spin-1 chains via $U_{j,\bar{j}}(\pi/\gamma)$ and then measuring the spin-1 sites \cite{Brennen2008,Miyake2010}. (b)~The gate fidelity of the $CZ$ gate for dynamically coupled spin-1 $XXZ$ chains with system size $L=20$ and $J,D\in[-6,6]$. ``Haldane" represents the Haldane phase protected by the $\mathbb Z_2\times \mathbb Z_2$ symmetry including the isotropic point $J=1,D=0$. ``FM'' and ``AFM'' denote the ferromagnetic and the antiferromagnetic phases, respectively. ``large-$D$'' denotes the trivial phase favored by the strong single-ion anisotropy $D \gg 0$.}
\label{fig01}
\end{figure}

In our previous work \cite{Ohta2025}, we found that the ground state of the spin-1 $XXZ$ model \cite{Chen2003,Kennedy1992} in the Haldane phase can achieve high gate fidelity of single-qubit gates by tuning the anisotropy. However, universal quantum computation requires two-qubit entangling gates, such as $CZ$ and $CNOT$ gates, in addition to single-qubit gates \cite{Barenco1995}. For the AKLT state, it has been shown that $CZ$ gates can be implemented by coupling two spin chains through a short dynamical interaction and then performing local measurements \cite{Brennen2008} (see Fig.~\hyperref[fig01]{1(a)}). Furthermore, Ref.~\citealp{Miyake2010} demonstrated that perfect single-qubit gates and a $CZ$ gate can be implemented throughout the isotropic Haldane phase by adiabatically decoupling individual spins from the bulk chains before performing local measurements. 
These constructions have so far been analyzed mainly for the AKLT state or isotropic resource states in the Haldane phase \cite{Brennen2008,Miyake2010}, and the question of how the anisotropy of the Hamiltonian affects the performance of two-qubit gates in the Haldane phase has remained open. 

In this Letter, we address this question for the 1D spin-1 $XXZ$ model with uniaxial single-ion anisotropy. As shown in Fig.~\hyperref[fig01]{1(a)}, we consider two spin-1 $XXZ$ chains dynamically coupled via a unitary gate as the resource state for the implementation of a $CZ$ gate and apply the same protocol as that used for the AKLT state \cite{Brennen2008}. We evaluate the gate fidelity \cite{Fujii2013, Masui2025} of a $CZ$ gate for the coupled spin-1 $XXZ$ chains and find that, while the gate fidelity for the isotropic AKLT state is $F_{CZ}=7/12\simeq0.583$, tuning the anisotropy within the Haldane phase toward the boundary with the antiferromagnetic (AFM) phase increases the fidelity significantly above the AKLT value (see Fig.~\hyperref[fig01]{1(b)}). We further derive a compact expression for the gate fidelity in terms of nonlocal correlation functions between the bulk and the chain ends, and the probability that the measurement fails, assuming only that the resource state of each chain lies in the Haldane phase. This expression reveals that AFM correlations enhance the gate fidelity of a $CZ$ gate, explaining its increase in the vicinity of the AFM phase. 
The realization of spin-1 chains with uniaxial anisotropy has been
proposed using dysprosium atoms and Rydberg platforms \cite{Brechtelsbauer2025,Mogerle2025}. Together with these proposals, our results suggest that the experimental realization of both single-qubit gates and two-qubit gates in MBQC could be possible using spin-1 chains with uniaxial anisotropy.

{\it The MBQC protocol for a $CZ$ gate.}---We first review the implementation of a $CZ$ gate using two AKLT chains based on the MBQC  protocol schematically illustrated in Fig.~\hyperref[fig01]{1(a)} \cite{Brennen2008}.
We consider two spin-1 chains in the AKLT state, each of which has $L$ spin-1s and two spin-$1/2$s attached at the ends, where the two spin-$1/2$s serve as the input and output qubits. The AKLT state can be created from arrayed spin-singlet pairs, by projecting adjacent spin-$1/2$s in each neighboring spin-singlet pairs onto the spin-1 subspace \cite{Affleck1987,Darmawan2010}.

To implement a $CZ$ gate, the $j$th spin-1s on chains A and B $(1\le j\le L)$ are dynamically coupled by the unitary evolution $U_{j,\bar{j}}(t)=\exp[-iH^{\rm int}_{j,\bar{j}}t]$ under the Hamiltonian $H^{\rm int}_{j,\bar{j}}=\gamma|+\rangle_{j}\langle{+}|\otimes|+\rangle_{\bar{j}}\langle{+}|$, where $|+\rangle,|\tilde{0}\rangle,|-\rangle$ are eigenstates of the $z$ component of the spin-1 operator $\bm S =(S^x,S^y,S^z)$, with eigenvalues $+1$, $0$, and $-1$, respectively (as shown in Fig.~\hyperref[fig01]{1(a)}, unbarred and barred site indices refer to chains A and B, respectively). The coupling between chains A and B is switched on at $t=0$ and switched off at $t=\pi/\gamma$, yielding the unitary operation
$U_{j,\bar{j}}(\pi/\gamma)=I_{j,\bar{j}}-2|+\rangle_{j}\langle{+}|\otimes|+\rangle_{\bar{j}}\langle{+}|$, 
where $I_{j,\bar{j}}$ is the identity operator acting on the $j$th and $\bar{j}$th sites.
Then, all spin-1s are simultaneously measured in the basis $\{|x\rangle,|y\rangle,|z\rangle\}$, where $|x\rangle=({-|+\rangle+|-\rangle})/{\sqrt{2}}$, $|y\rangle=({|+\rangle+|-\rangle})/{\sqrt{2}}$, and $|z\rangle=|\tilde{0}\rangle$. 

If the measurement outcomes on the $j$th and $\bar{j}$th sites are in either $x$ or $y$, the state of the spin-$1/2$s after the measurement can be expressed as 
\[
|\psi_{CZ}\rangle=CZ_{L+1,\overline{L+1}}|\phi^+\rangle_{0,L+1}|\phi^+\rangle_{\bar{0},\overline{L+1}},
\]
up to Pauli byproduct operators acting on the output spin-$1/2$s, where $|\phi^+\rangle=({|\uparrow\uparrow\rangle+|\downarrow\downarrow\rangle})/{\sqrt{2}}$ is a Bell state.
Otherwise, the same procedure yields the identity gate up to Pauli byproduct operators. Since the AKLT state and the measurement basis are both isotropic, the measurement outcomes are uniformly distributed and the $CZ$ gate is successfully implemented with probability $4/9$. 

In the case that $|\psi_{CZ}\rangle$ is successfully obtained, $|\chi^*\rangle$ and $|\omega^*\rangle$ are projected onto the $0$th spin-$1/2$s on chains A and B, respectively. This teleports the input states $|\chi\rangle$ and $|\omega\rangle$ to the output spin-$1/2$s at $L+1$th sites on chains A and B, respectively, with a $CZ$ gate applied. Although the output states generally acquire Pauli byproduct operators depending on the measurement outcomes on spin-1 sites, these operators can be corrected by applying the corresponding Pauli gates to the output spin-$1/2$s\cite{Brennen2008}. 

The state after the measurements on spin-1s and the correction of the Pauli byproducts can be written as
\begin{align}
\rho_{CZ}=\frac{4}{9}|\psi_{CZ}\rangle\langle{\psi_{CZ}}|
+\frac{5}{9}\sum_{a,b=\pm}p_{ab}|\phi^a\phi^b\rangle\langle{\phi^a\phi^b}|,
\label{eq01}
\end{align}
where $|\phi^a\phi^b\rangle\equiv|\phi^a\rangle_{0,L+1}|\phi^b\rangle_{\bar{0},\overline{L+1}}~(a,b=\pm)$ and $|\phi^{\pm}\rangle=({|\uparrow\uparrow\rangle\pm|\downarrow\downarrow\rangle})/{\sqrt{2}}$. Here, $p_{ab}$ is the conditional probability of obtaining $|\phi^a\phi^b\rangle$ when the $CZ$ gate implementation fails.
To quantify the potential as a resource for implementation of a $CZ$ gate, we evaluate the gate fidelity for a $CZ$ gate \cite{Fujii2013, Masui2025},
\begin{align}
F_{CZ}=\Tr_{0,L+1}\left[\Tr_{\bar{0},\overline{L+1}}\left[\rho_{CZ}|\psi_{CZ}\rangle\langle{\psi_{CZ}}|\right]\right].
\label{eq02}
\end{align}
Substituting Eq.~(\ref{eq01}) into Eq.~(\ref{eq02}), we obtain $F_{CZ}=7/12\simeq0.583$ for the AKLT state.

{\it Gate fidelity for spin-1 $XXZ$ chains.}---In the rest of this Letter, we apply the above protocol for implementation of a $CZ$ gate to two spin-1 $XXZ$ chains within the Haldane phase and evaluate the gate fidelity defined in Eq.~(\ref{eq02}) for dynamically coupled spin-1 $XXZ$ chains.

The Hamiltonian of the 1D spin-1 $XXZ$ model with uniaxial single-ion anisotropy \cite{Chen2003,Kennedy1992} whose ends are coupled with two spin-$1/2$s is given by \cite{Ohta2025}
\begin{align}
&H_{XXZ}(J,D)
=s^x_{0}{S^x_1}+s^y_{0}{S^y_1}+Js^z_{0}{S^z_1}\notag\\*
&+\sum_{i=1}^{L-1}\left({S^x_i}{S^x_{i+1}}+{S^y_i}{S^y_{i+1}}+J{S^z_i}{S^z_{i+1}}\right)+D\sum_{i=1}^L{(S^z_i)^2}\notag\\*
&+S^x_{L}s^x_{L+1}+S^y_{L}s^y_{L+1}+JS^z_{L}s^z_{L+1},
\label{eq03}
\end{align}
where $J$ and $D$ control the Ising-like anisotropy and the single-ion anisotropy, respectively, along the $z$ axis, and $\bm{s}=(X/2,Y/2,Z/2)$ is the spin-$1/2$ operator ($X$, $Y$, and $Z$ are the Pauli matrices). The first and last three terms in Eq.~(\ref{eq03}) represent the coupling between spin-1 and spin-$1/2$ at the left and right ends. 

We now consider implementing a $CZ$ gate using dynamically coupled spin-1 $XXZ$ chains within the Haldane phase. In the initial state, two spin-1 $XXZ$ chains are set in the ground state $|G(J,D)\rangle_A\otimes|G(J,D)\rangle_B$, where $|G(J,D)\rangle$ denotes the ground state of Eq.~(\ref{eq03}).
We now follow the protocol in Fig.~\hyperref[fig01]{1(a)}: the $j$th and $\bar{j}$th spin-1s on chains A and B are dynamically coupled by $U_{j,\bar{j}}(\pi/\gamma)$, then measurements of spin-1s are performed.
\begin{figure}[t]
\centering
\includegraphics[width=1.0\columnwidth]{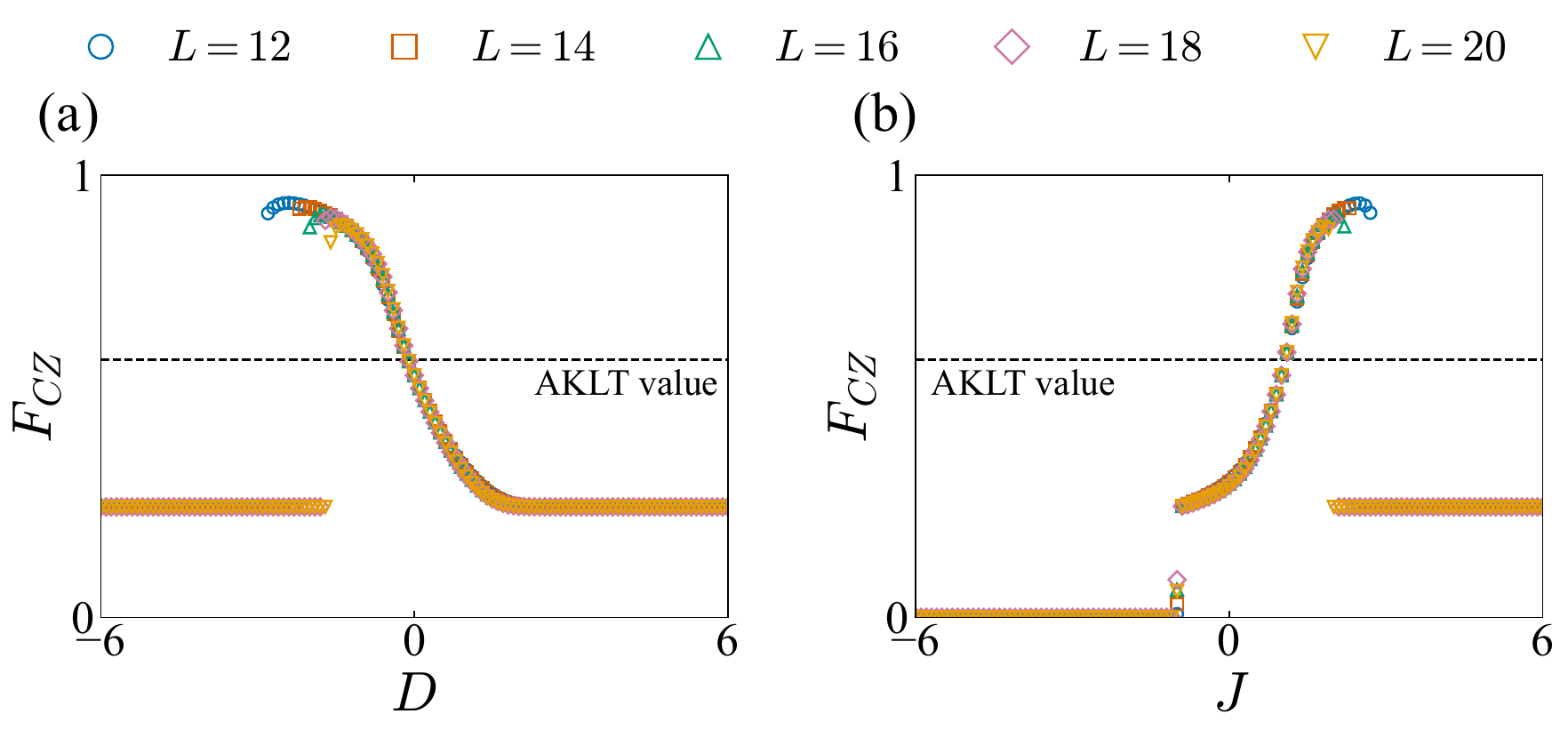}
\caption{Gate fidelity of a $CZ$ gate in Eq.~(\ref{eq02}) for the resource state $U_{j,\bar{j}}(\pi/\gamma)\left[|G(J,D)\rangle_A\otimes|G(J,D)\rangle_B\right]$ with chain lengths $12\le L \le 20$, where the $j$th and $\bar{j}$th spin-1s at the center of the chains are coupled by the unitary operation $U_{j,\bar{j}}(\pi/\gamma)$. (a) $J=1$ and $D\in[-6,6]$. (b) $D=0$ and $J\in[-6,6]$. The dashed lines indicate the gate fidelity for the AKLT state, $F_{CZ}=7/12$.} 
\label{fig02}
\end{figure}
\begin{figure}[t]
\centering
\includegraphics[width=1.0\columnwidth]{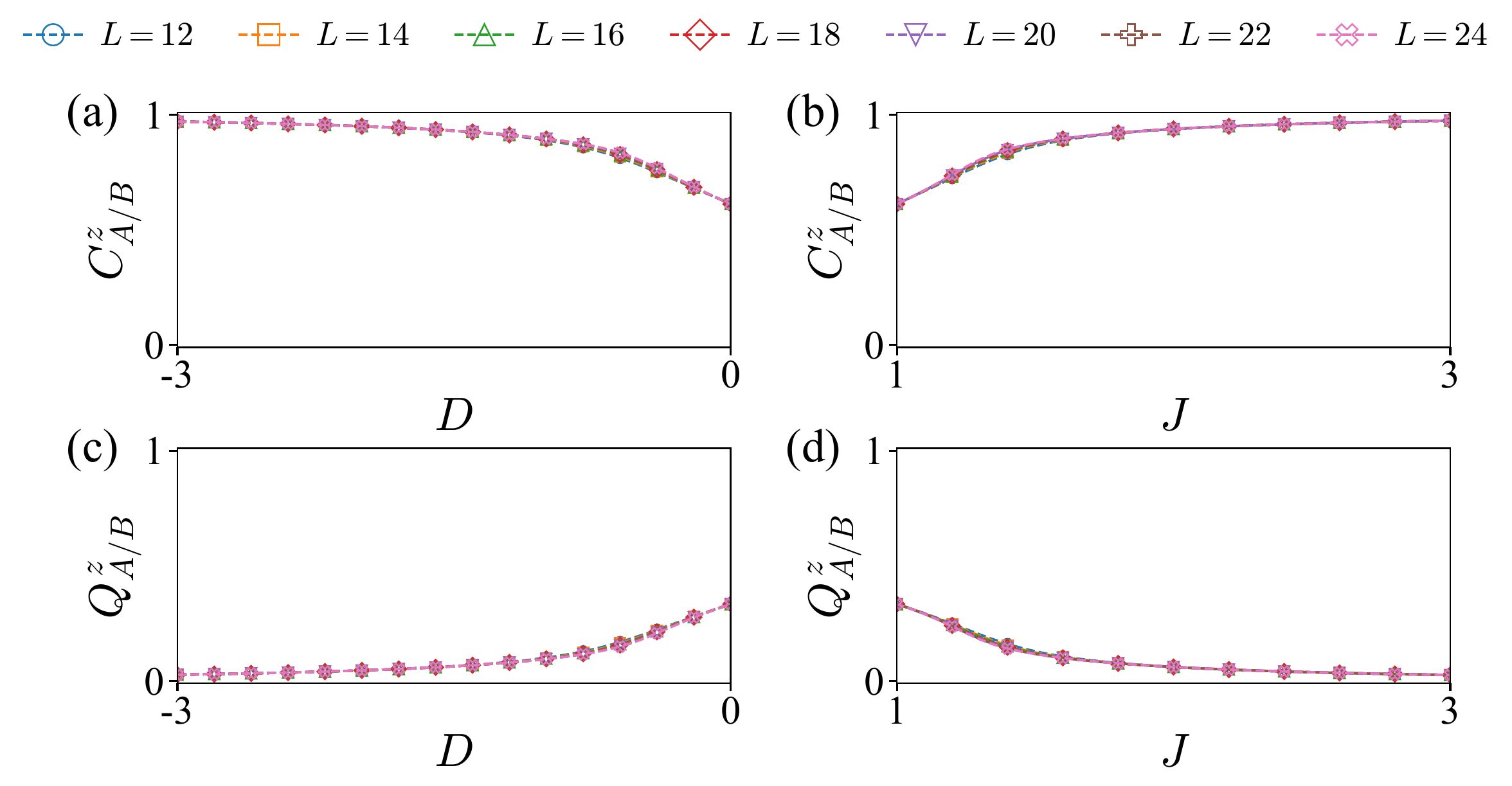}
\caption{(a),(b) Nonlocal correlation functions $C^z_{A/B}$ between the coupled site at the center of the chain and the output spin-$1/2$  and (c),(d) probabilities $Q^z_{A/B}$ of obtaining the outcome $z$, for the ground state of Eq.~(\ref{eq03}) with chain lengths $12\le L\le 24$. (a),(c) $J=1$ and $D\in[-3,0]$. (b),(d) $D=0$ and $J\in[1,3]$.}
\label{fig03}
\end{figure}
Figure~\hyperref[fig01]{1(b)} shows the gate fidelity of a $CZ$ gate for $J,D\in[-6,6]$ with $L=20$. Remarkably, the gate fidelity is enhanced in the presence of anisotropy close to the phase boundary between the Haldane phase and the AFM phase. To examine this enhancement in more detail, we show the gate fidelity of the $CZ$ gate for several chain lengths at $J=1$ and $D=0$ in Figs.~\hyperref[fig02]{2(a)} and \hyperref[fig02]{2(b)}, respectively.
The ground state $|G(J,D)\rangle$ is obtained using the density matrix renormalization group (DMRG) method \cite{White1992,SCHOLLWOCK2011} with a maximum bond dimension $400$ \cite{itensor}. As the single-ion anisotropy $D$ is decreased or the Ising-like anisotropy $J$ is increased along the $z$ axis, where the ground state approaches the AFM phase, the gate fidelity is remarkably enhanced beyond the AKLT value $F_{CZ}=7/12\simeq0.583$. Upon further decreasing $D$ or increasing $J$, the gate fidelity drops sharply in the vicinity of the AFM phase. 
Although the maximum gate fidelity decreases as the chain length $L$ increases, the finite-size extrapolation indicates that the maximum gate fidelity remains above the AKLT value $7/12$ in the thermodynamic limit~\cite{SM}.

This enhancement of the gate fidelity can be considered due to the nonlocal AFM correlations induced by tuning the anisotropy toward the AFM phase.
To see this, we derive a general expression for the gate fidelity of a $CZ$ gate assuming that the initial resource state $|\Psi\rangle_A\otimes|\Psi\rangle_B$, before the dynamical coupling, has the global $\mathbb{Z}_2\times\mathbb{Z}_2$ symmetry: \[
-\sigma^\mu_0\left(\prod_{k=1}^L e^{-i\pi S^\mu_k}\right)\sigma^\mu_{L+1}|\Psi\rangle=|\Psi\rangle~(\mu=x,y,z),
\]
where $\sigma^{\mu}$ denotes the $\mu$ component of the Pauli matrices.
Under this assumption, the gate fidelity of a $CZ$ gate can be expressed as \cite{SM}
\begin{align}
F_{CZ}=\frac14\left(1+C^z_A\right)\left(1+C^z_B\right)-\frac14\left(Q^z_AC^z_B+Q^z_BC^z_A\right).
\label{eq04}
\end{align}
Equation~(\ref{eq04}) shows that, as long as the $\mathbb{Z}_2\times\mathbb{Z}_2$ symmetry is preserved within the Haldane phase, the gate fidelity is determined by the nonlocal correlation functions $C^z_{A/B}$ and the failure probabilities $Q^z_{A/B}$. 

\begin{align*}
&C^z_A=-{}_A\langle{\Psi}|\left[S^z_j\left(\prod_{k=j+1}^Le^{-i\pi S^z_k}\right)Z_{L+1}\right]|\Psi\rangle_A,\notag\\
&C^z_B=-{}_B\langle{\Psi}|\left[S^z_{\bar{j}}\left(\prod_{\bar{k}=\overline{j+1}}^{\bar{L}}e^{-i\pi S^z_{\bar{k}}}\right)Z_{\overline{L+1}}\right]|\Psi\rangle_B\notag
\end{align*}
describe nonlocal correlations between the dynamically coupled site ($j$ and $\bar{j}$) and the output spin-$1/2$ of each chain.
\begin{align*}
Q^z_A=\lvert{}_A\langle{\Psi}|z\rangle_j\rvert^2,~Q^z_B=\lvert{}_B\langle{\Psi}|z\rangle_{\bar{j}}\rvert^2
\end{align*}
are the probabilities of obtaining the measurement outcome $z$ on the $j$th and $\bar{j}$th spin-1s, respectively, which makes implementation of a $CZ$ gate fail. We note that these expressions apply not only to the ground state of the
$XXZ$ model but also to {\it any} symmetry-preserving resource state in the Haldane phase.

Figures \hyperref[fig03]{3(a)--(d)} show $C^z_{A/B}$ and $Q^z_{A/B}$ for the ground state of Eq.~(\ref{eq03}). At the isotropic point $J=1$ and $D=0$, the probabilities of obtaining the measurement outcomes $x$, $y$, and $z$ are precisely equal. Consequently, $Q^z_{A/B}$ is close to $1/3$ and the gate fidelity is suppressed due to the large failure probability. As the single-ion anisotropy $D<0$ or the Ising-like anisotropy $J>1$ drives the ground state toward the AFM phase, the spin alignment is biased toward $|\pm\rangle$ states, which causes $C^z_{A/B}$ to approach unity and $Q^z_{A/B}$ to approach zero. 
The high gate fidelity in the spin-1 $XXZ$ chains therefore originates from the growth of AFM correlations near the AFM phase, which suppresses the failure outcomes.

{\it Conclusion and outlook.}---We have shown that dynamically coupled spin-1 $XXZ$ chains with uniaxial anisotropies in the Haldane phase can serve as a resource state for the implementation of a $CZ$ gate with high fidelity. Whereas the gate fidelity for the isotropic AKLT point is $7/12\simeq0.583$, the gate fidelity for the ground state of $XXZ$ chains is enhanced beyond the AKLT value when $D$ or $J$ is tuned toward the AFM phase. 
We have also derived a general expression for the gate fidelity of a $CZ$ gate assuming that the resource state of each chain lies in the Haldane phase and found that the gate fidelity is governed by nonlocal correlation functions between the coupled site and the output spin-$1/2$ and the failure probabilities.
Within the same framework, a $CNOT$ gate can be implemented analogously, and its gate fidelity exhibits the same qualitative dependence on the anisotropy parameters \cite{SM}.

Although the gate fidelities obtained here exceed the AKLT value, they remain below unity. It will therefore be important to investigate whether other models and interactions can produce higher fidelities. The realization of the spin-1 $XXZ$ chains with uniaxial anisotropies has been proposed in systems of dysprosium atoms confined in an optical lattice \cite{Brechtelsbauer2025}. Thus, developing a protocol for implementing two-qubit gates using dysprosium atoms is another important direction for future work.

{\it Acknowledgment.}---We thank D. Kagamihara and Y. Matsuzaki for helpful discussions. ST is supported by JSPS KAKENHI Grant No. 25K07158.

%

\clearpage
\onecolumngrid
\renewcommand{\combinedbibpart}{supplement}
\setcounter{section}{0}
\setcounter{subsection}{0}
\setcounter{subsubsection}{0}
\setcounter{equation}{0}
\setcounter{figure}{0}
\setcounter{table}{0}
\renewcommand{\thesection}{S\arabic{section}}
\renewcommand{\thesubsection}{\thesection.\arabic{subsection}}
\makeatletter
\renewcommand{\p@subsection}{}
\makeatother
\renewcommand{\theequation}{S\arabic{equation}}
\renewcommand{\thefigure}{S\arabic{figure}}
\renewcommand{\theHsection}{supp.\arabic{section}}
\renewcommand{\theHsubsection}{supp.\arabic{section}.\arabic{subsection}}
\renewcommand{\theHsubsubsection}{supp.\arabic{section}.\arabic{subsection}.\arabic{subsubsection}}
\renewcommand{\theHequation}{supp.\arabic{equation}}
\renewcommand{\theHfigure}{supp.\arabic{figure}}
\renewcommand{\theHtable}{supp.\arabic{table}}
\pdfbookmark[0]{Supplemental Material}{supplemental-material}
\begin{center}
{\large\bfseries Supplemental Material: ``High-Fidelity $CZ$ gate via Measurement-Based Quantum Computation Using Spin-1 \texorpdfstring{$XXZ$}{XXZ} Chains''\par}
\vspace{0.8em}
Hiroki Ohta and Shunji Tsuchiya\par
\vspace{0.3em}
{\small\itshape Department of Physics, Chuo University, Bunkyo, Tokyo 112-8551, Japan\par}
\end{center}

In this Supplemental Material, we provide the details of the calculations presented in the main text. In Sec.~\ref{sec:S1}, we introduce the matrix-product representation of the AKLT state and the global $\mathbb{Z}_2\times\mathbb{Z}_2$ symmetry operator used throughout this work. In Sec.~\ref{sec:S2}, we describe in detail the implementation of a $CZ$ gate using AKLT chains. In Sec.~\ref{sec:S3}, we derive a general expression for the gate fidelity of a $CZ$ gate in the presence of the global $\mathbb{Z}_2\times\mathbb{Z}_2$ symmetry. In Secs.~\ref{sec:S4} and \ref{sec:S5}, we describe the implementations of a $CNOT$ gate using AKLT chains and anisotropic spin-1 chains. Finally, in Sec.~\ref{sec:S6}, we examine the system-size dependence of the gate fidelity and extrapolate its maximum value to the thermodynamic limit.

\section{MATRIX PRODUCT REPRESENTATION OF THE AKLT STATE}
\label{sec:S1}
We now consider the two AKLT chains, given by
\begin{align}
|R\rangle=|{\rm AKLT}\rangle_A\otimes|{\rm AKLT}\rangle_B,
\label{eqS01}
\end{align}
where the AKLT state $|{\rm AKLT}\rangle$ is written as \cite{supp-Affleck1987,supp-Affleck1988}
\begin{align}
|{\rm AKLT}\rangle=\mathcal{N}
\left(\prod_{i=1}^{L} P_{i,i'} \right)\left[
|\psi^-\rangle_{0,1}
\left(\prod_{i=1}^{L-1} |\psi^-\rangle_{i',i+1}\right)
|\psi^-\rangle_{L',L+1}
\right],
\label{eqS02}
\end{align}
where $\mathcal N$ is the normalization factor. 
Here, $L$ is the number of spin-1s and the labels $0$ and $L+1$ denote the left and right ends of each chain, respectively (see Fig.~\ref{figS01}). $|\psi^-\rangle$ denotes a singlet pair of two spin-$1/2$s,
\begin{align}
|\psi^-\rangle=\frac{|\uparrow\downarrow\rangle-|\downarrow\uparrow\rangle}{\sqrt{2}},
\label{eqS03}
\end{align}
where $\ket{\uparrow}$ and $\ket{\downarrow}$ denote the eigenstates of the Pauli matrix $Z$ corresponding to the eigenvalues $+1$ and $-1$, respectively.
$P_{i,i'}$ projects two virtual spin-$1/2$s onto the physical spin-1 subspace as
\begin{align}
P_{i,i'}=
\frac{|+\rangle_i\langle{\uparrow\uparrow}|_{i,i'}}{\sqrt{3}}+\frac{|\tilde 0\rangle_i(\langle{\uparrow\downarrow}|+\langle{\downarrow\uparrow}|)_{i,i'}}{\sqrt{6}}+\frac{|-\rangle_i\langle{\downarrow\downarrow}|_{i,i'}}{\sqrt{3}},
\label{eqS04}
\end{align}
where $|+\rangle$, $|\tilde 0\rangle$, and $|-\rangle$ are the eigenstates of the $z$ component of the spin-1 operator $\bm S=(S^x,S^y,S^z)$ corresponding to the eigenvalues $+1$, $0$, and $-1$, respectively.
The AKLT state is the unique ground state of the AKLT Hamiltonian \cite{supp-Affleck1987,supp-Affleck1988,supp-Kennedy1992,supp-Brennen2008}
\begin{align}
H_{{\rm AKLT}}=&\bm{s}_{0}\cdot\bm{S}_1+\sum_{i=1}^{L-1}\left[\bm{S}_i\cdot\bm{S}_{i+1}+\frac{1}{3}(\bm{S}_i\cdot\bm{S}_{i+1})^2\right]+\bm{S}_{L}\cdot\bm{s}_{L+1},
\label{eqS05}
\end{align}
where $\bm{s}=(X/2,Y/2,Z/2)$ is the spin-$1/2$ operator ($X$, $Y$, and $Z$ are the Pauli matrices).

We use the matrix product state (MPS) formalism \cite{supp-Fannes1991,supp-Fannes1992,supp-Ostlund1995,supp-Perez2006,supp-SCHOLLWOCK2011,supp-Cirac2021}.
In this formalism, Eq.~(\ref{eqS02}) takes the form \cite{supp-Masui2025}
\begin{align}
|{\rm AKLT}\rangle=&\sum_{\sigma_1=0,1}\sum_{\sigma_2=0,1}\left(\prod_{i=1}^L\sum_{m_i=+,\tilde 0,-}\right)\langle{\sigma_1}|\Lambda\Gamma^{m_1}\Lambda\cdots\Lambda\Gamma^{m_L}\Lambda|\sigma_2\rangle\notag\\
&\times|\sigma_1\rangle_{0}|m_1\rangle_1\cdots|m_L\rangle_L|\sigma_2\rangle_{L+1},
\label{eqS06}
\end{align}
where the MPS tensors $\Lambda$ and $\Gamma^{m}$ represent a singlet bond and the projector onto the spin-1 subspace, respectively. They can be written as
\begin{align}
\Lambda=\frac{XZ}{\sqrt{2}},\ \Gamma^{\pm}=\frac{I\pm Z}{\sqrt{3}},\ \Gamma^{\tilde{0}}=\sqrt{\frac{2}{3}}X.
\label{eqS07}
\end{align}

The AKLT state possesses a $\mathbb Z_2\times \mathbb Z_2$ symmetry corresponding to global $\pi$ rotations about the $x$, $y$, and $z$ axes \cite{supp-Haldane1983,supp-Pollman2010,supp-Pollman2012,supp-Chen2013,supp-Wen2017}.
The global symmetry operators are defined as
\begin{align}
U^{\mu}=-\sigma^{\mu}_{\rm in}\left(\prod_{k=1}^Le^{-i\pi S^{\mu}_k}\right)\sigma^{\mu}_{\rm out},
\label{eqS08}
\end{align}
where $\mu=x, y, z$, and $\sigma^\mu$ denotes the $\mu$ component of the Pauli matrix.
The MPS tensors are transformed by $U^\mu$ as \cite{supp-Pollman2010,supp-Chen2011,supp-Shunch2011,supp-Cirac2021}
\begin{align}
\sum_{n}(e^{-i\pi S^\mu})_{mn}\Gamma^{n}\Lambda
=(\sigma^{\mu})^{\dagger}(\Gamma^{m}\Lambda)\sigma^{\mu}.
\label{eqS09}
\end{align}
Thus, $|{\rm AKLT}\rangle$ is invariant under $U^\mu$ as
\begin{align}
U^\mu|{\rm AKLT}\rangle=&-\sum_{\sigma_1=0,1}\sum_{\sigma_2=0,1}\left(\prod_{i=1}^L\sum_{m_i=+,\tilde 0,-}\right)\notag\\
&\times\langle{\sigma_1}|\sigma^{\mu}\Lambda(\sigma^{\mu})^{\dagger}\Gamma^{m_1}\Lambda\cdots\Lambda\Gamma^{m_L}\sigma^{\mu}(\sigma^{\mu})^{t}\Lambda|\sigma_2\rangle\notag\\
&\times|\sigma_1\rangle_{0}|m_1\rangle_1\cdots|m_L\rangle_L|\sigma_2\rangle_{L+1}\notag\\
=&|{\rm AKLT}\rangle,
\label{eqS10}
\end{align}
where $t$ denotes matrix transposition.
The same symmetry characterizes all states in the Haldane phase protected by $U^\mu$ \cite{supp-Haldane1983,supp-Pollman2010,supp-Pollman2012}. A general state $|\Psi\rangle$ in the Haldane phase is indeed invariant under $U^\mu$ as
\begin{align}
U^\mu|\Psi\rangle=|\Psi\rangle~(\mu=x,y,z).
\label{eqS11}
\end{align}

In the MPS description of MBQC, local measurements induce operations on the virtual degrees of freedom \cite{supp-GrossEisert2007,supp-GrossEtAl2007}.
To derive the post-measurement state below, it is convenient to rewrite Eq.~(\ref{eqS06}) in the following form \cite{supp-Brennen2008,supp-Miyake2010,supp-Barlett2010,supp-Darmawan2010}:
\begin{align}
|{\rm AKLT}\rangle=&\left(\prod_{i=1}^L\sum_{m_i=+,\tilde 0,-}\right)|m_1\rangle_1\cdots|m_L\rangle_L\notag\\
&\times\sqrt{2}\left( A^{m_L}_{L+1}A^{m_{L-1}}_{L+1}\dots A^{m_1}_{L+1}{\Lambda}_{L+1}\right)|\phi^+\rangle_{0,L+1}.
\label{eqS12}
\end{align}
Here, the operator $A^m$ is defined as $A^m = \Lambda\Gamma^m$, and $|\phi^+\rangle$ denotes the Bell state
\begin{align}
|\phi^+\rangle=\frac{|\uparrow\uparrow\rangle+|\downarrow\downarrow\rangle}{\sqrt{2}}.
\label{eqS13}
\end{align}
By using the following orthonormal basis
\begin{align}
|x\rangle=\dfrac{-|+\rangle+|-\rangle}{\sqrt{2}},
|y\rangle=\dfrac{|+\rangle+|-\rangle}{\sqrt{2}},
|z\rangle=|\tilde{0}\rangle,
\label{eqS14}
\end{align}
we can deform the representation in Eq.~(\ref{eqS12}) as
\begin{align}
|{\rm AKLT}\rangle=&\left(\prod_{i=1}^L\sum_{\mu_i=x,y,z}\right)|\mu_1\rangle_1\cdots|\mu_L\rangle_L\notag\\
&\times\sqrt{2}\left( B^{\mu_L}_{L+1}B^{\mu_{L-1}}_{L+1}\dots B^{\mu_1}_{L+1}\Lambda_{L+1}\right)|\phi^+\rangle_{0,L+1},
\label{eqS15}
\end{align}
where $B^{\mu}~(\mu = x,y,z)$ is defined as
\begin{align}
B^x = -\dfrac{X}{\sqrt{3}},
B^y = -\dfrac{iY}{\sqrt{3}},
B^z = -\dfrac{Z}{\sqrt{3}}.
\label{eqS16}
\end{align}
Thus, $|R\rangle$ in Eq.~(\ref{eqS01}) can be rewritten as
\begin{align}
|R\rangle=&\left(\prod_{i=1}^L\sum_{m_i=+,\tilde 0,-}\right)\left(\prod_{\bar{i}=\bar{1}}^{\bar{L}}\sum_{m_{\bar{i}}=+,\tilde 0,-}\right)|m_1\rangle_1\cdots|m_L\rangle_L|m_{\bar{1}}\rangle_{\bar{1}}\cdots|m_{\bar{L}}\rangle_{\bar{L}}\notag\\
&\times2\left( A^{m_L}_{L+1}A^{m_{\bar{L}}}_{\overline{L+1}}A^{m_{L-1}}_{L+1}A^{m_{\overline{L-1}}}_{\overline{L+1}}\dots A^{m_1}_{L+1}A^{m_{\bar{1}}}_{\overline{L+1}}\Lambda_{L+1}\Lambda_{\overline{L+1}}\right)|\phi^+\rangle_{0,L+1}|\phi^+\rangle_{\bar{0},\overline{L+1}}.
\label{eqS17}
\end{align}
\section{IMPLEMENTATION OF A $CZ$ GATE USING TWO COUPLED AKLT CHAINS}
\label{sec:S2}
\begin{figure}[t]
\begin{center}
\includegraphics[width=0.7\linewidth,clip]{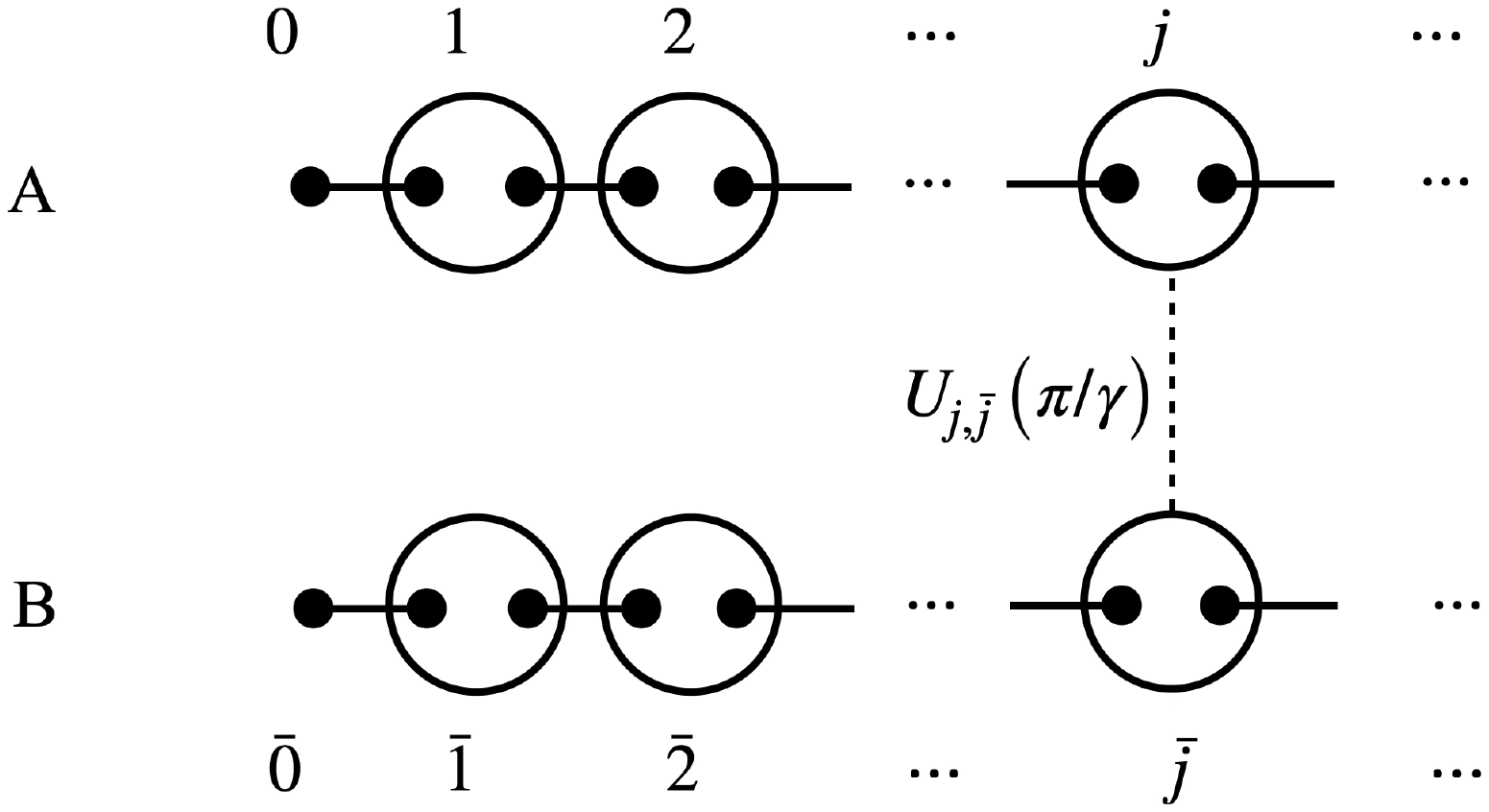}
\caption{Schematic illustration of the dynamically coupled two AKLT chains. The pair of black dots connected by a line represents a spin-singlet pair of spin-$1/2$s and the solid circles represent the spin-1 sites. The $j$th spin-1 on chain A and the $\bar{j}$th spin-1 on chain B are dynamically coupled  via the unitary operation in Eq.~(\ref{eqS20}).}
\label{figS01}
\end{center}
\end{figure}

We now consider implementing a $CZ$ gate using the AKLT chains as the resource state. First, the dynamical coupling between the $j$th and the $\bar{j}$th spin-1s $(1\le j\le L)$ on chains A and B is created via the unitary evolution
\begin{align}
U_{j,{\bar{j}}}(t)= {\rm exp}\left[{-iH^{\rm int}_{j,{\bar{j}}}t}\right],
\label{eqS18}
\end{align}
switched on during $0\le t \le \pi/\gamma$ under the Hamiltonian $H^{\rm int}_{j,{\bar{j}}}$ given by
\begin{align}
H^{\rm int}_{j,{\bar{j}}}=\gamma|+\rangle_{j}\langle{+}|\otimes|+\rangle_{\bar{j}}\langle{+}|
\label{eqS19}
\end{align}
(see Fig.~\ref{figS01} and Fig.~1 in the main text).
At $t=\pi/\gamma$, the above unitary can be represented as
\begin{align}
U_{j,{\bar{j}}}\left(\frac{\pi}{\gamma}\right)&=I_{j,\bar{j}}-\frac{1}{2}\left[S^z_{j}+\left(S^z_{j}\right)^2\right]\left[S^z_{\bar{j}}+\left(S^z_{\bar{j}}\right)^2\right]\notag\\
&=I_{j,\bar{j}}-2|+\rangle_{j}\langle{+}|\otimes|+\rangle_{\bar{j}}\langle{+}|,
\label{eqS20}
\end{align}
where $I_{j,\bar{j}}$ is the identity matrix for the $j$th and $\bar{j}$th sites.

Applying the unitary operation in Eq.~(\ref{eqS20}) to the two AKLT chains, we obtain
\begin{align}
U_{j,{\bar{j}}}\left(\frac{\pi}{\gamma}\right)|R\rangle=&|R\rangle-4\left(\prod_{i=1,i\ne j}^L\sum_{m_i=+,\tilde{0},-}\right)\left(\prod_{\bar{i}=\bar{1},\bar{i}\ne \bar{j}}^{\bar{L}}\sum_{m_{\bar{i}}=+,\tilde{0},-}\right)\notag\\
&\times|m_1\rangle\cdots|+\rangle_j\cdots|m_L\rangle|m_{\bar{1}}\rangle\cdots|+\rangle_{\bar{j}}\cdots|m_{\bar{L}}\rangle\notag\\
&\times\left[A^{m_L}_{L+1}\cdots\left(\dfrac{X_{L+1}Z_{L+1}+X_{L+1}}{\sqrt{6}}\right)\cdots A^{m_1}_{L+1}\Lambda_{L+1}\right]\notag\\
&\times\left[A^{m_{\bar{L}}}_{\overline{L+1}}\cdots\left(\dfrac{X_{\overline{L+1}}Z_{\overline{L+1}}+X_{\overline{L+1}}}{\sqrt{6}}\right)\cdots A^{m_{\bar{1}}}_{\overline{L+1}}\Lambda_{\overline{L+1}}\right]|\phi^+\rangle_{0,L+1}|\phi^+\rangle_{\bar{0},\overline{L+1}}.
\label{eqS21}
\end{align}
Using the basis defined in Eq.~(\ref{eqS14}), the above expression can be rewritten as
\begin{align}
|R\rangle-&2\left(\prod_{i=1,i\ne j}^L\sum_{\mu_i=x,y,z}\right)\left(\prod_{\bar{i}=\bar{1},\bar{i}\ne \bar{j}}^{\bar{L}}\sum_{\mu_{\bar{i}}=x,y,z}\right)\notag\\
&\times|\mu_1\rangle\cdots\left({-|x\rangle_j+|y\rangle_j}\right)\cdots|\mu_L\rangle|\mu_{\bar{1}}\rangle\cdots\left({-|x\rangle_{\bar{j}}+|y\rangle_{\bar{j}}}\right)\cdots|\mu_{\bar{L}}\rangle\notag\\
&\times\left[B^{\mu_L}_{L+1}\cdots\left(\dfrac{X_{L+1}Z_{L+1}+X_{L+1}}{\sqrt{6}}\right)\cdots B^{\mu_1}_{L+1}\Lambda_{L+1}\right]\notag\\
&\times\left[B^{\mu_{\bar{L}}}_{\overline{L+1}}\cdots\left(\dfrac{X_{\overline{L+1}}Z_{\overline{L+1}}+X_{\overline{L+1}}}{\sqrt{6}}\right)\cdots B^{\mu_{\bar{1}}}_{\overline{L+1}}\Lambda_{\overline{L+1}}\right]\notag\\
&\times|\phi^+\rangle_{0,L+1}|\phi^+\rangle_{\bar{0},\overline{L+1}}\notag\\
=&2\left(\prod_{i=1}^L\sum_{\mu_i=x,y,z}\right)\left(\prod_{\bar{i}=\bar{1}}^{\bar{L}}\sum_{\mu_{\bar{i}}=x,y,z}\right)|\mu_1\rangle\cdots|\mu_j\rangle_j\cdots|\mu_L\rangle|\mu_{\bar{1}}\rangle\cdots|\mu_{\bar{j}}\rangle_{\bar{j}}\cdots|\mu_{\bar{L}}\rangle\notag\\
&\times\left[B^{\mu_L}_{L+1}B^{\mu_{\bar{L}}}_{\overline{L+1}}\cdots C^{{\mu_j},{\mu_{\bar{j}}}}_{L+1,{\overline{L+1}}}\cdots B^{\mu_1}_{L+1}B^{\mu_{\bar{1}}}_{\overline{L+1}}\Lambda_{L+1}\Lambda_{\overline{L+1}}\right]|\phi^+\rangle_{0,L+1}|\phi^+\rangle_{\bar{0},\overline{L+1}}.
\label{eqS22}
\end{align}
Here, $C^{\mu,\nu}_{L+1,{\overline{L+1}}}~(\mu,\nu=x,y,z)$ is defined as
\begin{align}
C^{x,x}_{L+1,\overline{L+1}}
&=\dfrac{CZ_{L+1,\overline{L+1}}X_{L+1}X_{\overline{L+1}}}{3},
C^{x,y}_{L+1,\overline{L+1}}
=\dfrac{iCZ_{L+1,\overline{L+1}}X_{L+1}Y_{\overline{L+1}}}{3},\notag\\
C^{x,z}_{L+1,\overline{L+1}}
&=\dfrac{X_{L+1}Z_{\overline{L+1}}}{3},
C^{y,x}_{L+1,\overline{L+1}}
=\dfrac{iCZ_{L+1,\overline{L+1}}Y_{L+1}X_{\overline{L+1}}}{3},\notag\\
C^{y,y}_{L+1,\overline{L+1}}
&=-\dfrac{CZ_{L+1,\overline{L+1}}Y_{L+1}Y_{\overline{L+1}}}{3},
C^{y,z}_{L+1,\overline{L+1}}
=\dfrac{iY_{L+1}Z_{\overline{L+1}}}{3},\notag\\
C^{z,x}_{L+1,\overline{L+1}}
&=\dfrac{Z_{L+1}X_{\overline{L+1}}}{3},
C^{z,y}_{L+1,\overline{L+1}}
=\dfrac{iZ_{L+1}Y_{\overline{L+1}}}{3},
C^{z,z}_{L+1,\overline{L+1}}
=\dfrac{Z_{L+1}Z_{\overline{L+1}}}{3},
\label{eqS23}
\end{align}
where the $CZ$ gate is defined as $CZ_{L+1,{\overline{L+1}}}=I_{L+1,{\overline{L+1}}}-2|1\rangle_{L+1}\langle{1}|\otimes|1\rangle_{\overline{L+1}}\langle{1}|$.

We now assume that spin-1s are measured in the basis defined in Eq.~(\ref{eqS14}) and we obtain measurement outcomes
\begin{align}
\bm{m}=(\mu_1,\mu_{\overline{1}},\ldots,\mu_j,\mu_{\overline{j}},\ldots,\mu_{L},\mu_{\overline{L}}).
\label{eqS24}
\end{align}
Then, the state after all measurements, up to normalization, can be written as
\begin{align}
\left[B^{\mu_L}_{L+1}B^{\mu_{\bar{L}}}_{\overline{L+1}}\cdots C^{{\mu_j},{\mu_{\bar{j}}}}_{L+1,{\overline{L+1}}}\cdots B^{\mu_1}_{L+1}B^{\mu_{\bar{1}}}_{\overline{L+1}}\Lambda_{L+1}\Lambda_{\overline{L+1}}\right]|\phi^+\rangle_{0,L+1}|\phi^+\rangle_{\bar{0},\overline{L+1}}.
\label{eqS25}
\end{align}
From Eq.~(\ref{eqS23}), when the outcomes are $\mu_j,\mu_{\bar{j}}\in \{x,y\}$, the above state can be represented as
\begin{align}
\left[M^{(\bm{m})}_{L+1,{\overline{L+1}}}Y_{L+1}Y_{\overline{L+1}}\right]CZ_{L+1,{\overline{L+1}}}|\phi^+\rangle_{0,L+1}|\phi^+\rangle_{\bar{0},\overline{L+1}},
\label{eqS26}
\end{align}
where $M^{(\bm{m})}_{L+1,{\overline{L+1}}}$ represents the byproducts corresponding to a sequence of measurement outcomes $\bm{m}$.
Since the $CZ$ gate is a Clifford gate and the byproducts $M^{(\bm{m})}_{L+1,{\overline{L+1}}}$ and $Y_{L+1}Y_{\overline{L+1}}$ are the products of Pauli matrices, we can eliminate such byproducts via local operations and classical communication (LOCC) \cite{supp-Bennet1993,supp-Raussendorf2001,supp-Brennen2008,supp-Nielsen2011}. Therefore, we obtain the post-MBQC state as
\begin{align}
|\psi_{CZ}\rangle_{0,{\overline{0}},L+1,{\overline{L+1}}}=CZ_{L+1,{\overline{L+1}}}|\phi^+\rangle_{0,L+1}|\phi^+\rangle_{\bar{0},\overline{L+1}}.
\label{eqS27}
\end{align}
Projecting $|\chi^*\rangle$ and $|\omega^*\rangle$ onto the $0$th and $\bar{0}$th spin-$1/2$s, respectively, the input states $|\chi\rangle$ and $|\omega\rangle$ are teleported to the output qubits ($L+1$ and $\overline{L+1}$th spin-$1/2$s, respectively) with a $CZ$ gate applied:
\begin{align}
\langle{\chi^*}|_{0}\langle{\omega^*}|_{\bar{0}}|\psi_{CZ}\rangle_{0,{\overline{0}},L+1,{\overline{L+1}}}\propto CZ_{L+1,{\overline{L+1}}}|\chi\rangle_{L+1}|\omega\rangle_{\overline{L+1}}.
\label{eqS28}
\end{align}
On the other hand, when the measurement outcomes on the $j$th and $\bar{j}$th spin-1s include $z$, the state after all measurements can be written as
\begin{align}
\left[M^{(\bm{m})}_{L+1,{\overline{L+1}}}Y_{L+1}Y_{\overline{L+1}}\right]|\phi^+\rangle_{0,L+1}|\phi^+\rangle_{\bar{0},\overline{L+1}}.
\label{eqS29}
\end{align}
Therefore, after the elimination of the byproducts, the above state can be represented as $|\phi^+\rangle_{0,L+1}|\phi^+\rangle_{\bar{0},\overline{L+1}}$. For such a post-MBQC state, even if input states are supplied, the output states are simply the input states with the identity gate applied. Since the AKLT state is isotropic as well as the basis in Eq.~(\ref{eqS14}), the measurement outcomes are uniformly distributed and we can implement a $CZ$ gate with probability $4/9$ \cite{supp-Brennen2008}.
\section{DERIVATION OF THE GATE FIDELITY}
\label{sec:S3}
\subsection{Definition of the gate fidelity}
\label{subsec:S3A}
We define the gate fidelity \cite{supp-Fujii2013,supp-Masui2025} for a $CZ$ gate as
\begin{align}
F_{CZ}=\Tr_{0,L+1}\left[\Tr_{\bar{0},\overline{L+1}}\left[\rho_{CZ}|\psi_{CZ}\rangle_{0,{\overline{0}},L+1,{\overline{L+1}}}\langle{\psi_{CZ}}|\right]\right].
\label{eqS30}
\end{align}
Here, $\rho_{CZ}$ represents the post-MBQC state obtained when a state $\rho$ is used as the resource state and is given by
\begin{align}
\rho_{CZ} = \Tr_{\rm bulk}\left[\Tr_{\overline{\rm bulk}}\left[\sum_{\bm{m}}\mathcal{B}^{(\bm{m})}_{{L+1},\overline{L+1}} \mathcal{P}_{\bm{m}}\rho\mathcal{P}_{\bm{m}}\left(\mathcal{B}^{(\bm{m})}_{{L+1},\overline{L+1}} \right)^{\dagger}\right]\right],
\label{eqS31}
\end{align}
where $\Tr_{\rm bulk}$ and $\Tr_{\overline{\rm bulk}}$ are taken over the bulk spin-1s on chains A and B, respectively.
Here, the projection operators $\mathcal{P}_{\bm{m}}$ and the byproducts $\mathcal{B}^{(\bm{m})}_{{L+1},{\overline{L+1}}}$ are defined as
\begin{align}
\mathcal{P}_{\bm{m}}=&\left(\prod_{i=1}^{L}|\mu_{i}\rangle_i\langle{\mu_{i}}|\right)\otimes\left(\prod_{\bar{i}=\bar{1}}^{\bar{L}}|\mu_{\bar{i}}\rangle_{\bar{i}}\langle{\mu_{\bar{i}}}|\right),
\label{eqS32}\\
\mathcal{B}^{(\bm{m})}_{{L+1},{\overline{L+1}}}=&X_{L+1}X_{\overline{L+1}}\left(\prod_{i=L}^{1}\sigma^{\mu_i}_{L+1}\right)\left(\prod_{\bar{i}=\bar{L}}^{\overline{1}}\sigma^{\mu_{\bar{i}}}_{\overline{L+1}}\right)\left[\prod_{i=j}^{1}Z^{(1-\delta_{\mu_i,z})}_{\overline{L+1}}\right]\left[\prod_{\bar{i}=\bar{j}}^{\overline{1}}Z^{(1-\delta_{\mu_{\bar{i}},z})}_{L+1}\right]\notag\\
=&X^{N_x(\bm{A})+1}_{L+1}Z^{N_z{(\bm{A})}+N_x{(\tilde{\bm{B}}_{\bar{j}})}}_{L+1}X^{N_x(\bm{B})+1}_{\overline{L+1}}Z^{N_x{(\tilde{\bm{A}}_j)}+N_z{(\bm{B})}}_{\overline{L+1}},
\label{eqS33}
\end{align}
where $\bm{A}~(\bm{B})$ represents measurement outcomes on chain A (B), $\tilde{\bm{A}}_j~(\tilde{\bm{B}}_{\bar{j}})$ represents measurement outcomes up to the $j$ (${\bar{j}}$)th spin-1, and $N_x(~\cdot~)$ and $N_z(~\cdot~)$ denote the numbers of $X$ and $Z$ byproducts, respectively.

We now employ the stabilizer form for the post-MBQC state $|\psi_{CZ}\rangle$ as \cite{supp-Fujii2013,supp-Masui2025,supp-Nielsen2011}
\begin{align}
|\psi_{CZ}\rangle_{0,\bar{0},L+1,\overline{L+1}}\langle{\psi_{CZ}}|=&CZ_{L+1,\overline{L+1}}\left(|\phi^+\rangle_{0,L+1}|\phi^+\rangle_{\bar{0},\overline{L+1}}\langle{\phi^+}|_{0,L+1}\langle{\phi^+}|_{\bar{0},\overline{L+1}}\right)CZ_{L+1,\overline{L+1}}^{\dagger}\notag\\
=&CZ_{L+1,\overline{L+1}}\left(\frac{I_{0,L+1}+X_{0}X_{L+1}}{2}\right)\left(\frac{I_{0,L+1}+Z_{0}Z_{L+1}}{2}\right)\notag\\
&~~~\times\left(\frac{I_{\bar{0},\overline{L+1}}+X_{\bar{0}}X_{\overline{L+1}}}{2}\right)\left(\frac{I_{\bar{0},\overline{L+1}}+Z_{\bar{0}}Z_{\overline{L+1}}}{2}\right)CZ_{L+1,\overline{L+1}}^{\dagger}\notag\\
=&\left(\frac{I_{0,\bar{0},L+1,\overline{L+1}}+X_{0}X_{L+1}Z_{\overline{L+1}}}{2}\right)\left(\frac{I_{0,L+1}+Z_{0}Z_{L+1}}{2}\right)\notag\\
&\times~~~\left(\frac{I_{0,\bar{0},L+1,\overline{L+1}}+X_{\bar{0}}Z_{L+1}X_{\overline{L+1}}}{2}\right)\left(\frac{I_{\bar{0},\overline{L+1}}+Z_{\bar{0}}Z_{\overline{L+1}}}{2}\right).
\label{eqS34}
\end{align}
Using the relation \cite{supp-Masui2025}
\begin{align}
e^{-i\pi S^{\mu}}|\nu\rangle\langle{\nu}|=(-1)^{1-\delta_{\mu,\nu}}|\nu\rangle\langle{\nu}|,
\label{eqS35}
\end{align}
we can expand Eq.~(\ref{eqS30}) as
\begin{align}
F_{CZ}=\frac{1}{16}+\frac{1}{16}\sum_{i=1}^{15}F_i.
\label{eqS36}
\end{align}
Here, $F_{i}~(i=1,\ldots,15)$ denote
\begin{align}
F_1=&\Tr_{\rm all}\left[\rho X_{0}X_{\bar{0}}\left(\prod_{i=1}^j e^{-i\pi S^x_i}\right)\left(\prod_{i=j+1}^{L}e^{-i\pi S^y_i}\right)\right.\notag\\
&\left.~~~~~~~~~~~~~~~\times
\left(\prod_{\bar{i}=\bar{1}}^{\bar{j}}e^{-i\pi S^x_{\bar{i}}}\right)\left(\prod_{\bar{i}=\overline{j+1}}^{\bar{L}}e^{-i\pi S^y_{\bar{i}}}\right)Y_{L+1}Y_{\overline{L+1}}\right],\label{eqS37}\\
F_2=&\Tr_{\rm all}\left[\rho Y_{0}Y_{\bar{0}}\left(\prod_{i=1}^j e^{-i\pi S^y_i}\right)\left(\prod_{i=j+1}^{L}e^{-i\pi S^x_i}\right)\right.\notag\\
&\left.~~~~~~~~~~~~~~~\times
\left(\prod_{\bar{i}=\bar{1}}^{\bar{j}}e^{-i\pi S^y_{\bar{i}}}\right)\left(\prod_{\bar{i}=\overline{j+1}}^{\bar{L}}e^{-i\pi S^x_{\bar{i}}}\right)X_{L+1}X_{\overline{L+1}}\right],\label{eqS38}\\
F_3=&\Tr_{\rm all}\left[\rho Z_{0}Z_{\bar{0}}\left(\prod_{i=1}^L e^{-i\pi S^z_i}\right)\left(\prod_{\bar{i}=\bar{1}}^{\bar{L}}e^{-i\pi S^z_{\bar{i}}}\right)Z_{L+1}Z_{\overline{L+1}}\right],\label{eqS39}\\
F_4=&-\Tr_{\rm all}\left[\rho Z_{0}\left(\prod_{i=1}^L e^{-i\pi S^z_i}\right)Z_{L+1}\right],\label{eqS40}\\
F_5=&-\Tr_{\rm all}\left[\rho Z_{\bar{0}}\left(\prod_{\bar{i}=\bar{1}}^{\bar{L}} e^{-i\pi S^z_{\bar{i}}}\right)Z_{\overline{L+1}}\right],\label{eqS41}\\
F_6=&\Tr_{\rm all}\left[\rho X_{0}Z_{\bar{0}}\left(\prod_{i=1}^L e^{-i\pi S^x_i}\right)\left(\prod_{\bar{i}={\bar{1}}}^{\bar{j}}e^{-i\pi S^z_{\bar{i}}}\right)X_{L+1}\right],\label{eqS42}\\
F_7=&\Tr_{\rm all}\left[\rho Z_{0}X_{\bar{0}}\left(\prod_{i=1}^{j} e^{-i\pi S^z_i}\right)\left(\prod_{\bar{i}={\bar{1}}}^{\bar{L}}e^{-i\pi S^x_{\bar{i}}}\right)X_{\overline{L+1}}\right],\label{eqS43}\\
F_8=&-\Tr_{\rm all}\left[\rho X_{0}\left(\prod_{i=1}^L e^{-i\pi S^x_i}\right)\left(\prod_{\bar{i}={\overline{j+1}}}^{\bar{L}}e^{-i\pi S^z_{\bar{i}}}\right)X_{L+1}Z_{\overline{L+1}}\right],\label{eqS44}\\
F_9=&-\Tr_{\rm all}\left[\rho X_{\bar{0}}\left(\prod_{i=j+1}^L e^{-i\pi S^z_i}\right)\left(\prod_{\bar{i}={\bar{1}}}^{\bar{L}}e^{-i\pi S^x_{\bar{i}}}\right)Z_{L+1}X_{\overline{L+1}}\right],\label{eqS45}\\
F_{10}=&-\Tr_{\rm all}\left[\rho X_{0}Y_{\bar{0}}\left(\prod_{i=1}^j e^{-i\pi S^x_i}\right)\left(\prod_{i=j+1}^{L}e^{-i\pi S^y_i}\right)\right.\notag\\
&\left.~~~~~~~~~~~~~~~\times
\left(\prod_{\bar{i}=\bar{1}}^{\bar{j}}e^{-i\pi S^y_{\bar{i}}}\right)\left(\prod_{\bar{i}=\overline{j+1}}^{\bar{L}}e^{-i \pi S^x_{\bar{i}}}\right)Y_{L+1}X_{\overline{L+1}}\right],\label{eqS46}\\
F_{11}=&-\Tr_{\rm all}\left[\rho Y_{0}X_{\bar{0}}\left(\prod_{i=1}^j e^{-i\pi S^y_i}\right)\left(\prod_{i=j+1}^{L}e^{-i\pi S^x_i}\right)\right.\notag\\
&\left.~~~~~~~~~~~~~~~\times
\left(\prod_{\bar{i}=\bar{1}}^{\bar{j}}e^{-i\pi S^x_{\bar{i}}}\right)\left(\prod_{\bar{i}=\overline{j+1}}^{\bar{L}}e^{-i \pi S^y_{\bar{i}}}\right)X_{L+1}Y_{\overline{L+1}}\right],\label{eqS47}\\
F_{12}=&\Tr_{\rm all}\left[\rho Y_{0}Z_{\bar{0}}\left(\prod_{i=1}^L e^{-i\pi S^y_i}\right)\left(\prod_{\bar{i}={\bar{1}}}^{\bar{j}}e^{-i\pi S^z_{\bar{i}}}\right)Y_{L+1}\right],\label{eqS48}\\
F_{13}=&\Tr_{\rm all}\left[\rho Z_{0}Y_{\bar{0}}\left(\prod_{i=1}^j e^{-i\pi S^z_i}\right)\left(\prod_{\bar{i}={\bar{1}}}^{\bar{L}}e^{-i\pi S^y_{\bar{i}}}\right)Y_{\overline{L+1}}\right],\label{eqS49}\\
F_{14}=&-\Tr_{\rm all}\left[\rho Y_{0}\left(\prod_{i=1}^L e^{-i\pi S^y_i}\right)\left(\prod_{\bar{i}={\overline{j+1}}}^{\bar{L}}e^{-i\pi S^z_{\bar{i}}}\right)Y_{L+1}Z_{\overline{L+1}}\right],\label{eqS50}\\
F_{15}=&-\Tr_{\rm all}\left[\rho Y_{\bar{0}}\left(\prod_{i=j+1}^L e^{-i\pi S^z_i}\right)\left(\prod_{\bar{i}={\bar{1}}}^{\bar{L}}e^{-i\pi S^y_{\bar{i}}}\right)Z_{L+1}Y_{\overline{L+1}}\right],\label{eqS51}
\end{align}
where $\Tr_{\rm all}$ denotes taking trace over the entire system including $2L$ spin-1s and four spin-$1/2$s at the ends of the two chains.
Here, we use the completeness relation
\begin{align}
\sum_{\bm{m}}\mathcal{P}_{\bm{m}}=I_{\rm all},
\label{eqS52}
\end{align}
where $I_{\rm all}$ represents the identity matrix for the entire system.

As an example, we now consider the dynamically coupled AKLT chains in Eq.~(\ref{eqS21}) as a resource and evaluate the gate fidelity for the $CZ$ gate. As discussed in the previous section, a $CZ$ gate can be implemented successfully with probability $4/9$ and failure cases occur with probability $5/9$ which results in the implementation of the identity gate accompanied by the corresponding byproducts. The resource state after the process of MBQC is given by
\begin{align}
\rho_{CZ}
=&\frac{4}{9}|\psi_{CZ}\rangle_{0,\bar{0},L+1,\overline{L+1}}\langle{\psi_{CZ}}|+\frac{5}{9}\sum_{a,b=\pm}p_{ab}|\phi^a\phi^b\rangle\langle{\phi^a\phi^b}|,
\label{eqS53}
\end{align}
where $|\phi^a\phi^b\rangle\equiv|\phi^a\rangle_{0,L+1}|\phi^b\rangle_{\bar{0},\overline{L+1}}$ and $|\phi^{\pm}\rangle=({|\uparrow\uparrow\rangle\pm|\downarrow\downarrow\rangle})/{\sqrt{2}}$. Here, $p_{ab}$ are the conditional probabilities of obtaining $|\phi^a\phi^b\rangle$ for failure outcomes and the correction of byproducts in Eq.~(\ref{eqS33}).
Substituting $\rho_{CZ}$ and $|\psi_{CZ}\rangle_{0,\bar{0},L+1,\overline{L+1}}$ into Eq.~(\ref{eqS30}), we obtain the gate fidelity $F_{CZ}=7/12\simeq 0.583$.
\subsection{Reduction using the global symmetry}
\label{subsec:S3B}
We now derive a compact expression of the gate fidelity for a $CZ$ gate in the Haldane phase. As a preliminary step, we note that the unitary operation in Eq.~(\ref{eqS20}) acts on $u^{\mu}=e^{-i\pi S^{\mu}}~(\mu=x,y,z)$ as
\begin{align}
U_{j,\bar{j}}(\pi/\gamma)u^x_jU_{j,\bar{j}}(\pi/\gamma)=&u^x_j+2|-+\rangle_{j,\bar{j}}\langle{++}|+2|++\rangle_{j,\bar{j}}\langle{-+}|,\label{eqS54}\\
U_{j,\bar{j}}(\pi/\gamma)u^y_jU_{j,\bar{j}}(\pi/\gamma)=&u^y_j-2|-+\rangle_{j,\bar{j}}\langle{++}|-2|++\rangle_{j,\bar{j}}\langle{-+}|,\label{eqS55}\\
U_{j,\bar{j}}(\pi/\gamma)u^z_jU_{j,\bar{j}}(\pi/\gamma)=&u^z_j.\label{eqS56}
\end{align}
Denoting the states of chains A and B by $|\Psi\rangle_A$ and $|\Psi\rangle_B$, the resource state reads
\begin{align}
U_{j,\bar{j}}(\pi/\gamma)|\Psi\rangle_A\otimes|\Psi\rangle_B.
\label{eqS57}
\end{align}
We assume that the states of chains A and B are invariant under the global $\mathbb Z_2\times \mathbb Z_2$ symmetry as
\begin{align}
U^{\mu}_A|\Psi\rangle_A=|\Psi\rangle_A,~U^{\mu}_B|\Psi\rangle_B=|\Psi\rangle_B~(\mu=x,y,z).
\label{eqS58}
\end{align}
Now we focus on $F_1$ in Eq.~(\ref{eqS37}). It can be expanded as
\begin{align}
F_1=&\Tr\left[U_{j,\bar{j}}(\pi/\gamma)\left(|\Psi\rangle_A\langle{\Psi}|\otimes|\Psi\rangle_B\langle{\Psi}|\right)U_{j,\bar{j}}(\pi/\gamma)\left[X_0\left(\prod_{i=1}^ju^x_i\right)\right.\right.\notag\\
&\qquad\left.\left.\times\left(\prod_{i=j+1}^Lu^y_i\right)Y_{L+1}\right]\left[X_{\bar{0}}\left(\prod_{\bar{i}=\bar{1}}^{\bar{j}}u^x_{\bar{i}}\right)\left(\prod_{\bar{i}=\overline{j+1}}^{\bar{L}}u^y_{\bar{i}}\right)Y_{\overline{L+1}}\right]\right]\notag\\
=&\Tr\left[|\Psi\rangle_A\langle{\Psi}|\left[X_0\left(\prod_{i=1}^ju^x_i\right)\left(\prod_{i=j+1}^Lu^y_i\right)Y_{L+1}\right]\right]\notag\\
&\times\Tr\left[|\Psi\rangle_B\langle{\Psi}|\left[X_{\bar{0}}\left(\prod_{\bar{i}=\bar{1}}^{\bar{j}}u^x_{\bar{i}}\right)\left(\prod_{\bar{i}=\overline{j+1}}^{\bar{L}}u^y_{\bar{i}}\right)Y_{\overline{L+1}}\right]\right]\notag\\
&-2\Tr\left[|\Psi\rangle_A\langle{\Psi}|\left[X_0\left(\prod_{i=1}^{j-1}u^x_i\right)|-\rangle_j\langle{+}|\left(\prod_{i=j+1}^Lu^y_i\right)Y_{L+1}\right]\right]\notag\\
&\times\Tr\left[|\Psi\rangle_B\langle{\Psi}|\left[X_{\bar{0}}\left(\prod_{\bar{i}=\bar{1}}^{\overline{j-1}}u^x_{\bar{i}}\right)|-\rangle_{\bar{j}}\langle{+}|\left(\prod_{\bar{i}=\overline{j+1}}^{\bar{L}}u^y_{\bar{i}}\right)Y_{\overline{L+1}}\right]\right]\notag\\
&-2\Tr\left[|\Psi\rangle_A\langle{\Psi}|\left[X_0\left(\prod_{i=1}^{j-1}u^x_i\right)|+\rangle_j\langle{-}|\left(\prod_{i=j+1}^Lu^y_i\right)Y_{L+1}\right]\right]\notag\\
&\times\Tr\left[|\Psi\rangle_B\langle{\Psi}|\left[X_{\bar{0}}\left(\prod_{\bar{i}=\bar{1}}^{\overline{j-1}}u^x_{\bar{i}}\right)|+\rangle_{\bar{j}}\langle{-}|\left(\prod_{\bar{i}=\overline{j+1}}^{\bar{L}}u^y_{\bar{i}}\right)Y_{\overline{L+1}}\right]\right].
\label{eqS59}
\end{align}
For an arbitrary unitary matrix $V$ which anticommutes with $U^{\mu}$ in Eq.~(\ref{eqS08}), $\langle{\Psi}|V|\Psi\rangle=0$ holds, if $|\Psi\rangle$ has $\mathbb Z_2\times \mathbb Z_2$ symmetry. Therefore, Eq.~(\ref{eqS59}) can be rewritten as
\begin{align}
F_1 = &2\Tr\left[|\Psi\rangle_A\langle{\Psi}||+\rangle_j\langle{+}|\left(\prod_{i=j+1}^Lu^z_i\right)Z_{L+1}\right]\Tr\left[|\Psi\rangle_B\langle{\Psi}||+\rangle_{\bar{j}}\langle{+}|\left(\prod_{\bar{i}=\overline{j+1}}^{\bar{L}}u^z_{\bar{i}}\right)Z_{\overline{L+1}}\right]\notag\\
&+2\Tr\left[|\Psi\rangle_A\langle{\Psi}||-\rangle_j\langle{-}|\left(\prod_{i=j+1}^Lu^z_i\right)Z_{L+1}\right]\Tr\left[|\Psi\rangle_B\langle{\Psi}||-\rangle_{\bar{j}}\langle{-}|\left(\prod_{\bar{i}=\overline{j+1}}^{\bar{L}}u^z_{\bar{i}}\right)Z_{\overline{L+1}}\right].
\label{eqS60}
\end{align}
Applying a similar manipulation to other terms in Eq.~(\ref{eqS36}), we can rewrite Eq.~(\ref{eqS36}) as
\begin{align}
F_{CZ}=&\frac{1}{4}+\frac{1}{2}\Tr\left[\left(|\Psi\rangle_A\langle{\Psi}|\otimes|\Psi\rangle_B\langle{\Psi}|\right)\left(|++\rangle_{j,\bar{j}}\langle{++}|+|--\rangle_{j,\bar{j}}\langle{--}|\right)\right.\notag\\*
&\qquad\left.\times\left(\prod_{i=j+1}^Lu^z_i\right)\left(\prod_{\bar{i}=\overline{j+1}}^{\bar{L}}u^z_{\bar{i}}\right)Z_{L+1}Z_{\overline{L+1}}\right]\notag\\
&-\frac12\left\{\Tr\left[\left(|\Psi\rangle_A\langle{\Psi}|\otimes|\Psi\rangle_B\langle{\Psi}|\right)\left(|++\rangle_{j,\bar{j}}\langle{++}|+|-+\rangle_{j,\bar{j}}\langle{-+}|\right)\left(\prod_{\bar{i}=\overline{j+1}}^{\bar{L}}u^z_{\bar{i}}\right)Z_{\overline{L+1}}\right]\right.\notag\\
&\left.+\Tr\left[\left(|\Psi\rangle_A\langle{\Psi}|\otimes|\Psi\rangle_B\langle{\Psi}|\right)\left(|++\rangle_{j,\bar{j}}\langle{++}|+|+-\rangle_{j,\bar{j}}\langle{+-}|\right)\left(\prod_{i=j+1}^Lu^z_i\right)Z_{L+1}\right]\right\}.
\label{eqS61}
\end{align}
For the following projection operators
\begin{align}
|+\rangle\langle{+}|=&\frac12\left(|x\rangle\langle{x}|+|y\rangle\langle{y}|-|x\rangle\langle{y}|-|y\rangle\langle{x}|\right),\label{eqS62}\\
|-\rangle\langle{-}|=&\frac12\left(|x\rangle\langle{x}|+|y\rangle\langle{y}|+|x\rangle\langle{y}|+|y\rangle\langle{x}|\right),\label{eqS63}
\end{align}
since $|x\rangle\langle{x}|$ and $|y\rangle\langle{y}|$ commute with $U^{\mu}$, we obtain
\begin{align}
F_{CZ}=\frac14\left(1+C^z_A\right)\left(1+C^z_B\right)-\frac14\left(Q^z_AC^z_B+Q^z_BC^z_A\right).
\label{eqS64}
\end{align}
Here,
\begin{align}
&C^z_A=-{}_A\langle{\Psi}|\left[S^z_j\left(\prod_{i=j+1}^L u^z_i\right)Z_{L+1}\right]|\Psi\rangle_A,\notag\\
&C^z_B=-{}_B\langle{\Psi}|\left[S^z_{\bar{j}}\left(\prod_{\bar{i}=\overline{j+1}}^{\bar{L}}u^z_{\bar{i}}\right)Z_{\overline{L+1}}\right]|\Psi\rangle_B
\label{eqS65}
\end{align}
represent the nonlocal correlation functions between the $j$th and $L+1$th ($\bar{j}$th and $\overline{L+1}$th) spins along the $z$ axis, and
\begin{align}
Q^z_A=\lvert{}_A\langle{\Psi}|z\rangle_j\rvert^2,~Q^z_B=\lvert{}_B\langle{\Psi}|z\rangle_{\bar{j}}\rvert^2
\label{eqS66}
\end{align}
represent the probabilities of obtaining the outcomes $z$ on the $j$th and $\bar{j}$th spins, respectively, resulting in the failure of a $CZ$ gate. The gate fidelity for a $CZ$ gate is thus determined by the nonlocal correlation functions and the failure probabilities in the Haldane phase.
\section{IMPLEMENTATION OF A $CNOT$ GATE}
\label{sec:S4}
Analogous to the implementation of a $CZ$ gate in Sec.~\ref{sec:S2}, we consider implementing a $CNOT$ gate using the dynamically coupled AKLT chains. In this case, the $j$th and the $\bar{j}$th spin-1s on chains A and B are dynamically coupled via the unitary evolution under the Hamiltonian 
\begin{align}
H^{\rm int}_{j,{\bar{j}}}=\gamma|+_z\rangle_{j}\langle{+_z}|\otimes|+_x\rangle_{\bar{j}}\langle{+_x}|,
\label{eqS67}
\end{align}
where $|+_x\rangle$ and $|+_z\rangle$ represent eigenstates of $S^x$ and $S^z$, respectively, each corresponding to the eigenvalue $+1$.
At $t=\pi/\gamma$, the unitary operator is
\begin{align}
U_{j,{\bar{j}}}\left(\frac{\pi}{\gamma}\right)=&I_{j,\bar{j}}-\frac{1}{2}\left[S^z_{j}+\left(S^z_{j}\right)^2\right]\left[S^x_{\bar{j}}+\left(S^x_{\bar{j}}\right)^2\right]\notag\\
=&I_{j,\bar{j}}-2|+_z\rangle_{j}\langle{+_z}|\otimes|+_x\rangle_{\bar{j}}\langle{+_x}|.
\label{eqS68}
\end{align}
The corresponding wave function is deformed as
\begin{align}
U_{j,{\bar{j}}}\left(\frac{\pi}{\gamma}\right)|R\rangle=&|R\rangle-4\left(\prod_{i=1,i\ne j}^L\sum_{m_i=+,\tilde{0},-}\right)\left(\prod_{\bar{i}=\bar{1},\bar{i}\ne \bar{j}}^{\bar{L}}\sum_{m_{\bar{i}}=+,\tilde{0},-}\right)\notag\\
&\times|m_1\rangle\cdots|+_z\rangle_j\cdots|m_L\rangle|m_{\bar{1}}\rangle\cdots|+_x\rangle_{\bar{j}}\cdots|m_{\bar{L}}\rangle\notag\\
&\times\left[A^{m_L}_{L+1}\cdots\left(\dfrac{X_{L+1}Z_{L+1}+X_{L+1}}{\sqrt{6}}\right)\cdots A^{m_1}_{L+1}\Lambda_{L+1}\right]\notag\\
&\times\left[A^{m_{\bar{L}}}_{\overline{L+1}}\cdots\left(\dfrac{X_{\overline{L+1}}Z_{\overline{L+1}}-Z_{\overline{L+1}}}{\sqrt{6}}\right)\cdots A^{m_{\bar{1}}}_{\overline{L+1}}\Lambda_{\overline{L+1}}\right]\notag\\
&\times|\phi^+\rangle_{0,L+1}|\phi^+\rangle_{\bar{0},\overline{L+1}}.
\label{eqS69}
\end{align}
Using the basis defined in Eq.~(\ref{eqS14}), the above expression can be rewritten in the same form as Eq.~(\ref{eqS22}), where $C^{\mu,\nu}_{L+1,{\overline{L+1}}}~(\mu,\nu=x,y,z)$ is defined as
\begin{align}
C^{x,x}_{L+1,\overline{L+1}}
&=\frac{X_{L+1}X_{\overline{L+1}}}{3},
C^{x,y}_{L+1,\overline{L+1}}
=\frac{i\,CNOT_{L+1,\overline{L+1}}X_{L+1}Y_{\overline{L+1}}}{3},\notag\\
C^{x,z}_{L+1,\overline{L+1}}
&=\frac{CNOT_{L+1,\overline{L+1}}X_{L+1}Z_{\overline{L+1}}}{3},
C^{y,x}_{L+1,\overline{L+1}}
=\frac{iY_{L+1}X_{\overline{L+1}}}{3},\notag\\
C^{y,y}_{L+1,\overline{L+1}}
&=-\frac{CNOT_{L+1,\overline{L+1}}Y_{L+1}Y_{\overline{L+1}}}{3},
C^{y,z}_{L+1,\overline{L+1}}
=\frac{i\,CNOT_{L+1,\overline{L+1}}Y_{L+1}Z_{\overline{L+1}}}{3},\notag\\
C^{z,x}_{L+1,\overline{L+1}}
&=\frac{Z_{L+1}X_{\overline{L+1}}}{3},
C^{z,y}_{L+1,\overline{L+1}}
=\frac{iZ_{L+1}Y_{\overline{L+1}}}{3},
C^{z,z}_{L+1,\overline{L+1}}
=\frac{Z_{L+1}Z_{\overline{L+1}}}{3},
\label{eqS70}
\end{align}
where the $CNOT$ gate is defined as $CNOT_{L+1,{\overline{L+1}}}=|0\rangle_{L+1}\langle{0}|\otimes I_{\overline{L+1}}+|1\rangle_{L+1}\langle{1}|\otimes X_{\overline{L+1}}$.

From Eq.~(\ref{eqS70}), when the outcomes are $\mu_j=x,y$ and $\mu_{\bar{j}}=y,z$, the state after all measurements can be represented as
\begin{align}
\left[M^{(\bm{m})}_{L+1,{\overline{L+1}}}Y_{L+1}Y_{\overline{L+1}}\right]CNOT_{L+1,{\overline{L+1}}}|\phi^+\rangle_{0,L+1}|\phi^+\rangle_{\bar{0},\overline{L+1}},
\label{eqS71}
\end{align}
where $M^{(\bm{m})}_{L+1,{\overline{L+1}}}$ represents the byproducts corresponding to a sequence of measurement outcomes $\bm{m}$.
Since the byproducts $M^{(\bm{m})}_{L+1,{\overline{L+1}}}$ and $Y_{L+1}Y_{\overline{L+1}}$ are the products of Pauli matrices, we can eliminate them via LOCC. Therefore, we obtain the post-MBQC state as
\begin{align}
|\psi_{CNOT}\rangle_{0,{\overline{0}},L+1,{\overline{L+1}}}=&CNOT_{L+1,{\overline{L+1}}}|\phi^+\rangle_{0,L+1}|\phi^+\rangle_{\bar{0},\overline{L+1}}.
\label{eqS72}
\end{align}
Otherwise, after the elimination of the byproducts, the state after all measurements becomes $|\phi^+\rangle_{0,L+1}|\phi^+\rangle_{\bar{0},\overline{L+1}}$, and the identity gate is implemented. Due to the isotropy, we can implement a $CNOT$ gate with probability $4/9$ similar to a $CZ$ gate.

The structure of Eq.~(\ref{eqS70}) is identical to that of Eq.~(\ref{eqS23}) up to the replacement of the measurement outcomes on the $\bar{j}$th spin-1 $\mu_{\bar{j}}=x,y$ by $\mu_{\bar{j}}=y,z$ that make the gate succeed. Thus, the discussion of the implementation of a $CZ$ gate in the preceding sections applies equally to a $CNOT$ gate by replacing the $z$ axis on chain B with the $x$ axis.
The definition of the gate fidelity for a $CNOT$ gate is therefore given by Eq.~(\ref{eqS30}) with $|\psi_{CZ}\rangle$ replaced by $|\psi_{CNOT}\rangle$, and the same manipulation as in Sec.~\ref{subsec:S3A} gives the expansion of the same form as Eq.~(\ref{eqS36}). Consequently, the gate fidelity is obtained as $F_{CNOT}=7/12\simeq 0.583$ for the coupled AKLT chains.
\section{$CNOT$ GATE USING ANISOTROPIC CHAINS}
\label{sec:S5}
We consider implementing a $CNOT$ gate using the two coupled anisotropic spin-1 chains as a resource state. Here, we evaluate the gate fidelity for a $CNOT$ gate with respect to the ground state of the following Hamiltonian \cite{supp-Chen2003,supp-Kennedy1992}
\begin{align}
H_{XXZ-XZZ}(J,D)=&H^A_{XXZ}(J,D)+H^B_{XZZ}(J,D)\notag\\*
=&s^x_{0}{S^x_1}+s^y_{0}{S^y_1}+Js^z_{0}{S^z_1}+\sum_{i=1}^{L-1}\left({S^x_i}{S^x_{i+1}}+{S^y_i}{S^y_{i+1}}+J{S^z_i}{S^z_{i+1}}\right)\notag\\*&+D\sum_{i=1}^L{(S^z_i)^2}
+S^x_{L}s^x_{L+1}+S^y_{L}s^y_{L+1}+JS^z_{L}s^z_{L+1}\notag\\
&+Js^x_{\bar{0}}{S^x_{\bar{1}}}+s^y_{\bar{0}}{S^y_{\bar{1}}}+s^z_{\bar{0}}{S^z_{\bar{1}}}+\sum_{\bar{i}=\bar{1}}^{\overline{L-1}}\left(J{S^x_{\bar{i}}}{S^x_{\overline{i+1}}}+{S^y_{\bar{i}}}{S^y_{\overline{i+1}}}+{S^z_{\bar{i}}}{S^z_{\overline{i+1}}}\right)\notag\\*
&+D\sum_{{\bar{i}}={{\bar{1}}}}^{\bar{L}}{(S^x_{\bar{i}})^2}+JS^x_{\bar{L}}s^x_{\overline{L+1}}+S^y_{\bar{L}}s^y_{\overline{L+1}}+S^z_{\bar{L}}s^z_{\overline{L+1}},
\label{eqS73}
\end{align}
where $H^A_{XXZ}(J,D)$ represents the $XXZ$ model with the anisotropies $J$ and $D$ along the $z$ axis, and $H^B_{XZZ}(J,D)$ represents the $XZZ$ model with the anisotropies along the $x$ axis. The ground state of the above Hamiltonian is represented as $|G(J,D)\rangle=|G(J,D)\rangle_A\otimes|G(J,D)\rangle_B$ where $|G(J,D)\rangle_A$ and $|G(J,D)\rangle_B$ are the ground states of the Hamiltonians $H^A_{XXZ}(J,D)$ and $H^B_{XZZ}(J,D)$, respectively. We take $U_{j,\bar{j}}(\pi/\gamma)|G(J,D)\rangle$ as the resource state for a $CNOT$ gate.

For a resource state satisfying Eq.~(\ref{eqS58}), the gate fidelity of a $CNOT$ gate is obtained by applying the same manipulation as in Sec.~\ref{subsec:S3B}:
\begin{align}
F_{CNOT}=\frac14\left(1+C^z_A\right)\left(1+C^x_B\right)-\frac14\left(Q^z_AC^x_B+Q^x_BC^z_A\right).
\label{eqS74}
\end{align}
Here,
\begin{align}
C^x_B=-{}_B\langle{\Psi}|\left[S^x_{\bar{j}}\left(\prod_{\bar{i}=\overline{j+1}}^{\bar{L}}u^x_{\bar{i}}\right)X_{\overline{L+1}}\right]|\Psi\rangle_B
\label{eqS75}
\end{align}
represents the nonlocal correlation function between the $\bar{j}$ and $\overline{L+1}$th spins along the $x$ axis, and
\begin{align}
Q^x_B=\lvert{}_B\langle{\Psi}|x\rangle_{\bar{j}}\rvert^2
\label{eqS76}
\end{align}
represents the probability of obtaining the outcome $x$ on the $\bar{j}$th spin, resulting in the failure of a $CNOT$ gate. The gate fidelity for a $CNOT$ gate is also determined by the nonlocal correlation function and the failure probability in the Haldane phase. Therefore, the gate fidelity for the ground state of Eq.~(\ref{eqS73}) gives similar behavior to that for a $CZ$ gate.
\section{SYSTEM SIZE DEPENDENCE OF THE GATE FIDELITY}
\label{sec:S6}
\begin{figure}[b]
\begin{center}
\includegraphics[width=\linewidth]{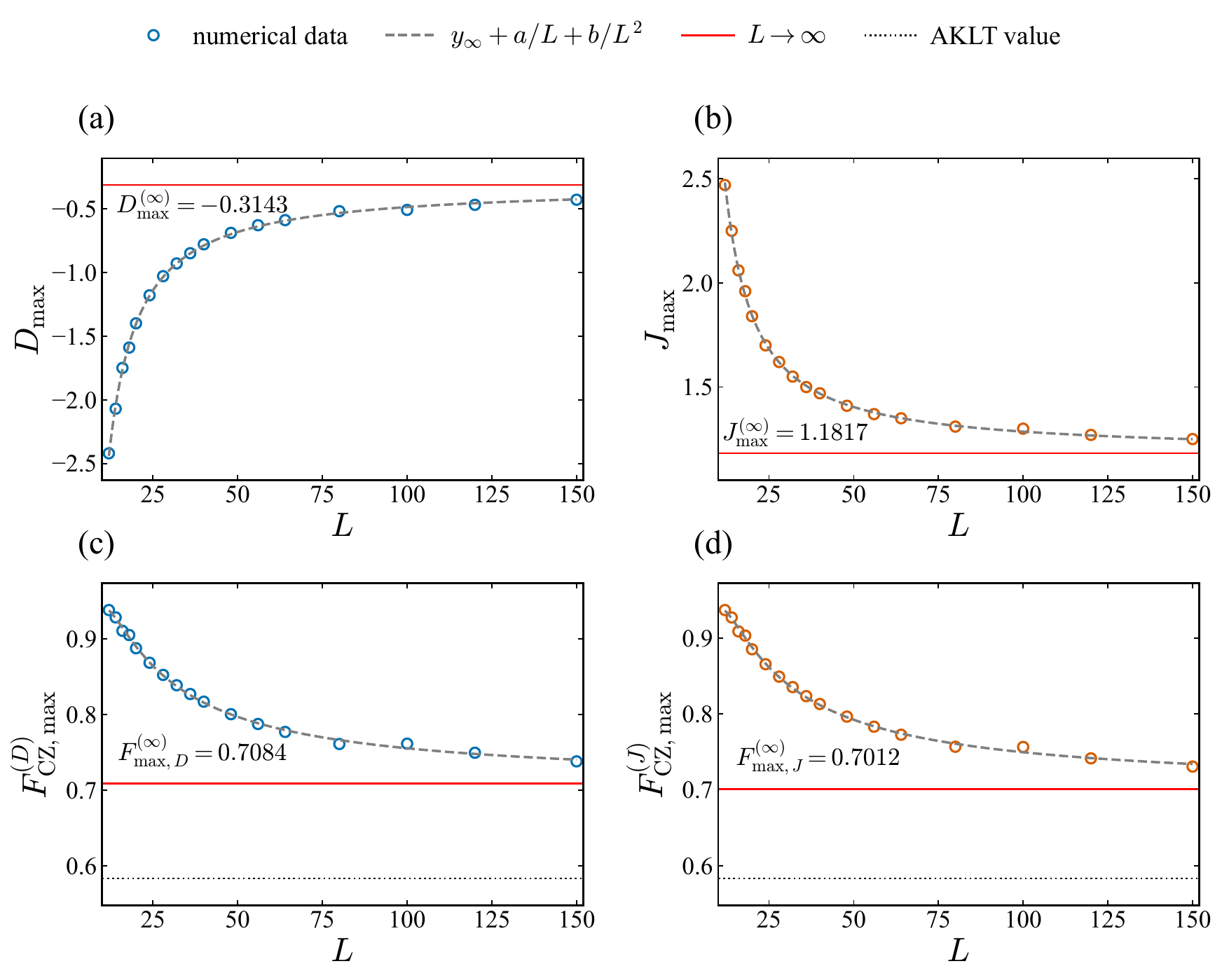}
\caption{System size dependence of (a) $D_{\rm max}$ and (b) $J_{\rm max}$ at which the gate fidelity of the $CZ$ gate takes its maximum, and of the maximum values of the gate fidelity (c) $F^{(D)}_{CZ,{\rm max}}$ for $J=1$ and (d) $F^{(J)}_{CZ,{\rm max}}$ for $D=0$, for $12\le L \le 150$. Open circles are the numerical data, dashed lines are the fits by $y_{\infty}+a/L+b/L^2$, and solid lines are the extrapolated values for $L\to\infty$. Dotted lines in (c) and (d) represent the AKLT value $F_{CZ}=7/12$. Ground states are obtained by the density matrix renormalization group method \cite{supp-White1992,supp-SCHOLLWOCK2011} with a maximum bond dimension of 400 {\cite{supp-itensor}}.}
\label{figS02}
\end{center}
\end{figure}

In Fig.~2 of the main text, we find that the gate fidelity of a $CZ$ gate depends on the chain length $L$, and the values of $D$ and $J$ at which the gate fidelity changes sharply depend on the system size. We here examine this dependence in detail. For each $L$, we denote by $D_{\rm max}$ and $J_{\rm max}$ the values of $D$ and $J$ at which the gate fidelity takes its maximum. We also denote by $F^{(D)}_{CZ,{\rm max}}$ and $F^{(J)}_{CZ,{\rm max}}$ the maximum values of the gate fidelity for $J=1$ and for $D=0$, respectively. Figures~\hyperref[figS02]{S2(a),~(b)} and \hyperref[figS02]{S2(c),~(d)} show $D_{\rm max}$ and $J_{\rm max}$, and  $F^{(D)}_{CZ,{\rm max}}$ and $F^{(J)}_{CZ,{\rm max}}$, respectively, for $L=12$--$150$.

We fit the numerical data in the form $y_{\infty}+a/L+b/L^2$, which gives
\begin{align}
D_{\rm max}(L)=&-0.3143-\frac{16.2030}{L}-\frac{111.8034}{L^2},
\label{eqS77}\\
J_{\rm max}(L)=&\ \ 1.1817+\frac{9.6953}{L}+\frac{70.9140}{L^2},
\label{eqS78}\\
F^{(D)}_{CZ,{\rm max}}(L)=&\ \ 0.7084+\frac{4.9460}{L}-\frac{26.3825}{L^2},
\label{eqS79}\\
F^{(J)}_{CZ,{\rm max}}(L)=&\ \ 0.7012+\frac{5.1112}{L}-\frac{27.4992}{L^2}.
\label{eqS80}
\end{align}
Taking $L\to\infty$, we obtain $D^{(\infty)}_{\rm max}=-0.3143,~F^{(\infty)}_{{\rm max},D}=0.7084$ and $J^{(\infty)}_{\rm max}=1.1817,~F^{(\infty)}_{{\rm max},J}=0.7012$ for the maximum values of the gate fidelity. Both $D^{(\infty)}_{\rm max}$ and $J^{(\infty)}_{\rm max}$ are finite, and the gate fidelity takes its maximum at $D<0$ and $J>1$ also in the thermodynamic limit. The extrapolated values, $F^{(\infty)}_{{\rm max},D}=0.7084$ and $F^{(\infty)}_{{\rm max},J}=0.7012$ are still beyond the AKLT value $F_{CZ}=7/12\simeq0.583$. This indicates that the enhancement of the gate fidelity by the anisotropy is not a finite-size effect and remains in the thermodynamic limit.
\end{document}